\documentclass[a4paper,11pt]{article}
\pdfoutput=1 

\usepackage{jinstpub} 

\usepackage{lineno}
\usepackage{enumitem}
\usepackage{textgreek}

\title{A laserball calibration device for the SNO+ scintillator phase}

\author[1,2]{S. Valder,\note{Corresponding author.}\note{Now at Rutherford Appleton Laboratory, Didcot, OX11 0DE, UK.}}
\author[]{A. Gibson-Foster,}
\author[]{E. Falk,}
\author[]{S.J.M. Peeters,}
\author[]{C. Mills,}
\author[]{M. Nirkko,}
\author[]{M. Rigan,}
\author[3]{and J. Sinclair\note{Now at SLAC National Accelerator Laboratory, Menlo Park, CA 94025, USA.}}

\affiliation[]{Department of Physics and Astronomy, University of Sussex,\\Pevensey II, Falmer, Brighton, BN1 9RH, UK}

\emailAdd{s.l.valder@sussex.ac.uk}
\emailAdd{sammy.valder@stfc.ac.uk}

\abstract{Located 2~km underground in SNOLAB, Sudbury, Canada, SNO+ is a large scale liquid scintillator experiment that primarily aims to search for neutrinoless double beta decay. Whilst SNO+ has light and radioactive calibration sources external to the inner volume, an internally deployed optical source is necessary for the full characterization of the detector model. A laser diffuser ball developed for SNO has previously demonstrated to be an effective optical calibration device for both SNO and SNO+ water phase. Since the introduction of liquid scintillator for SNO+, the material compatibility, cleanliness, and radiopurity requirements of any materials in contact with the internal medium have increased. Improving on the original SNO laserball design, a new laserball calibration device has been developed for the SNO+ scintillator phase with the goal of measuring the optical properties of the detector and performing routine PMT gain and timing calibrations. Simulations have been written to model the diffusion properties to optimise optical and temporal performance for calibration. Prototype laserballs have been built and characterised, demonstrating sub-ns timing resolution and a quasi-isotropic light distribution.}


\collaboration[c]{}

\begin{document}
\maketitle
\flushbottom

\section{Introduction}
\label{sec:intro}

An evolution of the Nobel prize winning Sudbury Neutrino Observatory (SNO) experiment~\cite{SNO}, SNO+~\cite{SNOplus} is a next generation multi-purpose neutrino experiment with a primary goal of detecting neutrinoless double beta decay~\cite{Hartnell_2012}. The SNO+ detector consists of a 6~m radius spherical acrylic vessel (AV) surrounded by a geodesic steel structure containing 9362~inward-facing Hamamatsu R1408 8-in photomultiplier tubes (PMTs) at an average distance of 8.35~m from the centre of the AV. Equipped with light concentrators, the PMTs have an effective optical coverage of approximately 54\%. Instead of D$_{2}$O, SNO+ uses poly(p-phenylene oxide) (PPO) doped linear alkylbenzene (LAB) as an active medium. In addition, bis-MSB and Te will be added to the liquid scintillator to increase light yield and act as a 0\textnu\textbeta\textbeta~source respectively~\cite{SNOplus}. A 6.8~m tall acrylic cylindrical neck with radius~0.75~m extends from the AV allowing calibration sources to be deployed~\cite{opticalwaterpaper}. External from the AV, 7000~tonnes of ultra-pure water is used to shield against radioactivity from detector components and the surrounding rock. A full description of the detector can be found in~\cite{SNO,SNOplus}.

In April 2022, the process of filling the AV with liquid scintillator and PPO wavelength shifter was completed. As the detector transitioned into the scintillator phase there was a need to update and re-evaluate calibration sources compatibility. The current laserball~\cite{rfordphdthesis,rfordmasters,opticalcalibrationhardwareSNO} has been used in both SNO and SNO+ water-phase~\cite{opticalwaterpaper} and will continue to be used for calibrations inside the external water region. However, the change in active mediums for SNO+ brings a new material compatibility requirement ensuring negligible chemical effect between the liquid scintillator and any material subject to internal deployment. Additionally, an increase in cleanliness and radiopurity demands for the SNO+ scintillator phase have made a replacement laserball necessary.

A new laserball calibration source has been developed primarily at the University of Sussex for the SNO+ scintillator phase building on the design of the original laserball. This paper will provide an overview of the device, the newly developed simulation, and the experimental tests and results. The work presented contains the research and development efforts that led to the shipping of a prototype laserball to SNOLAB, where the optimisation and commissioning of the final laserball source will occur. 

\section{Calibration Requirements} \label{sec:calibrequirements}

To achieve the SNO+ physics goals, a thorough \textit{in situ} understanding of the optical properties of the detector and detector medium is needed. These include the scintillator light yield, absorption and scattering coefficients, as well as PMT gains, time offsets and efficiencies~\cite{SNOplus}. SNO+ will take advantage of a number of complementary calibration sources including a new scintillator phase laserball~\cite{caden2020status}. The scintillator phase laserball has the capacity to perform routine PMT gain and timing calibrations as well as make measurements of the optical properties of the detector which can then be used to calibrate the detector simulation model~\cite{opticalwaterpaper}.

It is essential to have an accurate knowledge of the relative times at which PMTs detect the photons of a specific event. This attribute is directly correlated to the performance and systematic uncertainties associated with reconstruction algorithms including particle identification. SNO+ has a dedicated timing re-calibration system named TELLIE~\cite{alves2015calibration}. This helps to mitigate the number of internal calibration deployments which can carry an inherent contamination risk to the liquid scintillator. TELLIE uses LED light injection through optical fibres distributed around the detector to measure the timing response of the PMTs. However as TELLIE has multiple LED sources it requires previous calibration to establish relative offsets~\cite{michalthesis}. The scintillator phase laserball provides a single-source complementary PMT timing calibration which is more analytically straightforward, less frequent, and can be used as a reference analysis providing the offsets for TELLIE. The data required for PMT calibration analyses require high timing resolutions defined by the PMT temporal response. In SNO+, the Hamamatsu R1408 PMTs have a single-photoelectron timing resolution standard deviation below 1.7~ns~\cite{SNO}. In order to achieve PMT timing calibration, a temporal uncertainty of below 1.0~ns is set as a requirement for the scintillator phase laserball. 


As a single calibration source the laserball data can be used for measurements of the attenuation lengths of different detector mediums providing a greater understanding of the optical properties of the detector. The scintillator phase laserball will provide attenuation measurements from source to PMT through liquid scintillator, AV, and water; which can then be used in conjunction with measurements from water phase to extract the attenuation coefficients~\cite{opticalwaterpaper}. Moreover, the scintillator phase laserball can be used for PMT gain calibration and to measure the angular response of the PMTs and concentrators. These measurements require a well characterised light source that is ideally quasi-isotropic. The water phase laserball was designed with an initial uniformity of 10\% over a polar angle field of view of approximately $300^{\circ}$~\cite{rfordphdthesis, rfordmasters}, and an azimuthal intensity distribution consistent to 3\%~\cite{opticalcalibrationhardwareSNO, opticalwaterpaper, nunobarrosphdthesis}. These act as the loose requirements for the scintillator phase laserball. In reality any characterised anisotropy can be corrected for at the analysis level, as previously demonstrated in the optical calibration of SNO~\cite{moffatphdthesis,opticalcalibrationhardwareSNO}.

The change in active medium for SNO+ also brings new material compatibility and technical requirements for the laserball source. Any materials that could potentially come into contact with the liquid scintillator must pass material compatibility tests to ensure there is negligible chemical effect on both the LAB and subject material. Furthermore, there is a technical requirement to be able to determine the position and orientation of the source when deployed, independently from the side ropes attached to the carriage. To do this an additional light source, named the Umbilical Flasher Object (UFO), will sequentially pulse LEDs oriented in a circular formation. A wavelength of 660~nm is chosen so that the underwater camera system in SNO+ can observe it, whilst being at an acceptable quantum efficiency level (<1\%) of the PMTs~\cite{biller1999measurements}.

\section{Scintillator Phase Laserball}
\label{sec:laserball}

\begin{figure}[htbp]
\centering 
\includegraphics[width=.4\textwidth,origin=c]{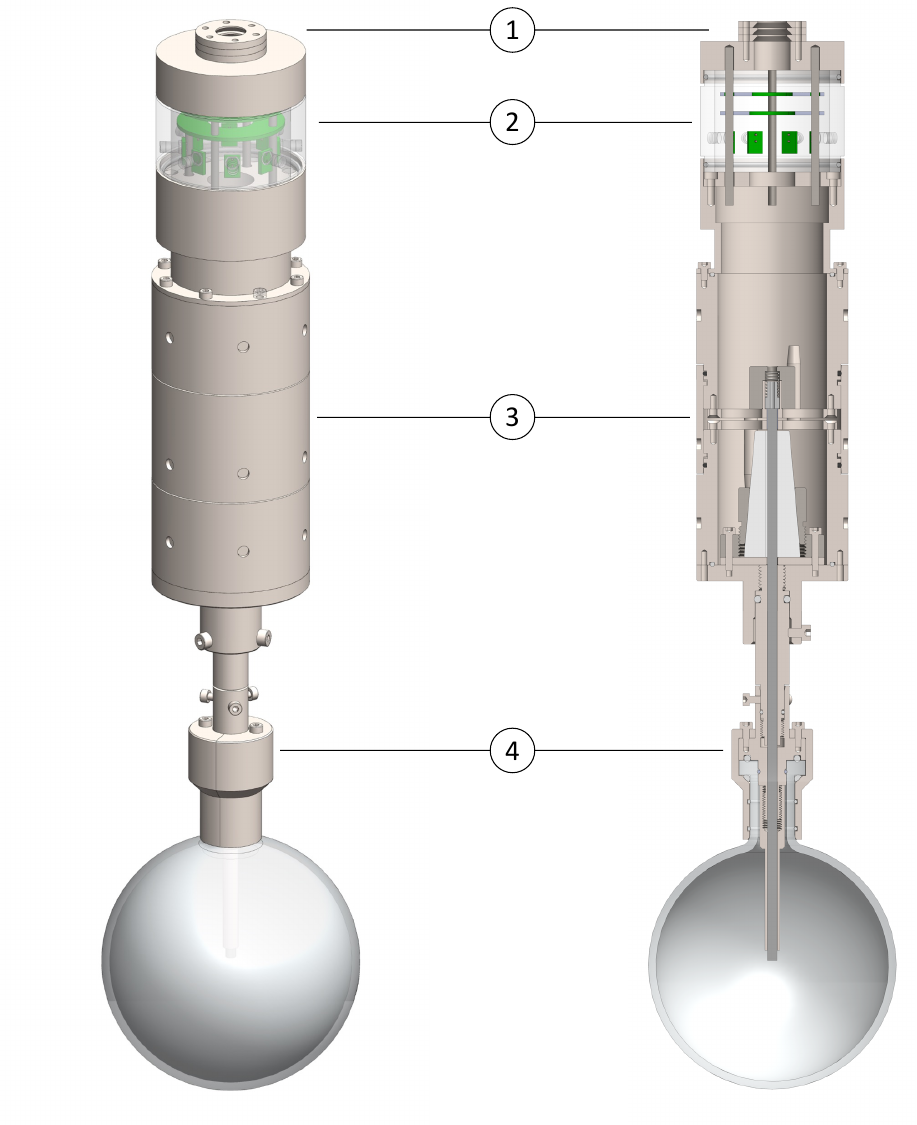}
\quad
\includegraphics[width=.168\textwidth]{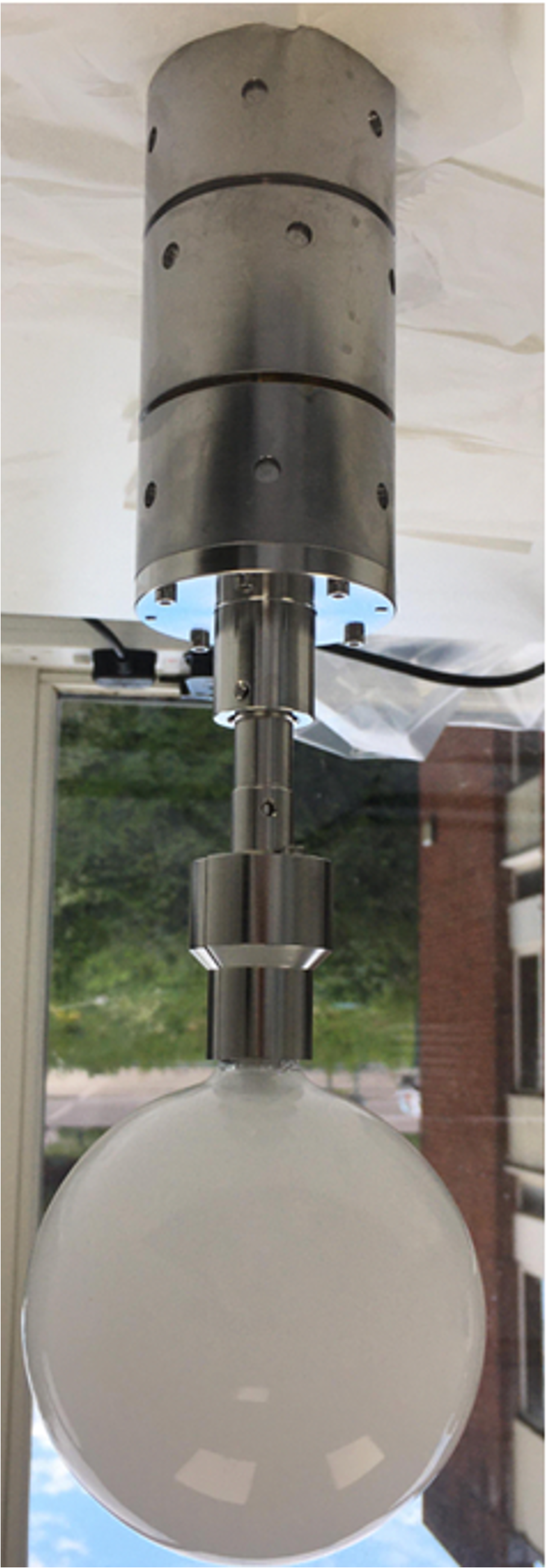}
\caption{\label{fig:laserball_CAD} The full scintillator phase source connector + laserball design, shown in 3D (left) and as a cross section (centre). The parts are labelled as follows: (1) The o-ring gasket stack for umbilical seal, (2) the Umbilical Flasher Object (UFO), (3) the source connector, and (4) the laserball calibration source. A picture of a laserball prototype is also shown (right).}
\end{figure}

The scintillator phase laserball design builds upon the original laserball inherited from SNO~\cite{SNO,opticalcalibrationhardwareSNO}. A full CAD model of the laserball and source connector assembly is shown in figure~\ref{fig:laserball_CAD}. Light is guided through a fibre bundle from the umbilical which connects to a rigid quartz rod inside the source connector, the rod penetrates into a spherical flask filled with glass microspheres suspended in silicone gel which act to diffuse the light. The design underwent three major improvements: a redesign of the neck to reduce self-shadowing effects, an adjustable central quartz rod to fine tune the light injection point as to ensure a quasi-isotropic light source, and a new source connector design enabling simple interchanges between different radioactive and optical sources. 

\subsection{Light injection}

The laserball light injection system is provided by a pulsed nitrogen laser (337~nm) which drives a selectable dye laser system~\cite{ford1998snolaser}. For scintillator phase, six wavelengths are available with well-characterised spectra centered at 337~nm, 385~nm, 405~nm, 420~nm, 450~nm, and 500~nm. The laser has an inherent typical operating rate between 10~Hz and 40~Hz. The laser system is coupled to an optical fibre bundle which guides the output light to the source connector through an umbilical cable. The stability of the laser emission can be monitored live during calibration campaigns by looking at the integrated number of PMT hits per second. Previous campaigns have demonstrated the fluctuations in number of PMTs registering hits per laserball pulse were smaller than 2\%~\cite{opticalwaterpaper}.

\subsection{Laserball assembly}

The full assembly, unless otherwise stated, is made out of electropolished 316L stainless steel. Injected into the source connector cylinder, the optical fibre bundle is glued inside a shroud which sits within a custom-made optical coupling. A 4~mm diameter quartz rod is coupled to the end of the fibre bundle through a spring which ensures the fibre and rod are always in contact. A cone-shaped PTFE clamp with a vertical opening is used to hold the rod in place whilst also protecting it from direct contact with any stainless steel components to avoid scratching or breaking when tightened. The clamp is wrench tightened using a bespoke bolt-ring design which closes the opening and secures the quartz rod with a friction fit. The source connector assembly is aligned using two alignment rods which pass through the inner plates.

\begin{figure}[htbp]
\centering 
\includegraphics[width=.5\textwidth,origin=c]{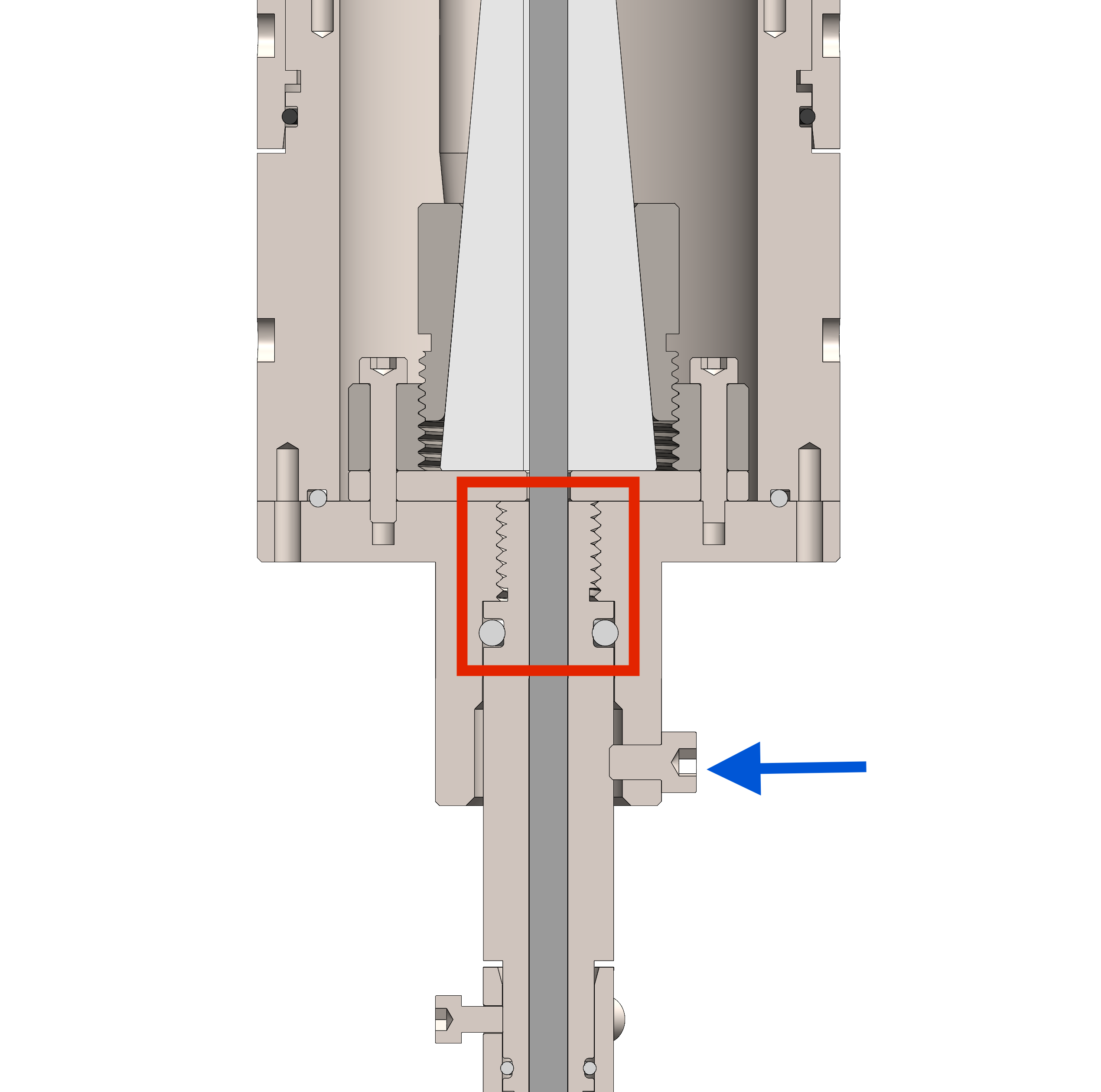}
\caption{\label{fig:adjustment} A close up of the mechanism used to fine tune the light injection point within the laserball flask. The screw fit between the source connector plate and the neck tube is shown in the red box. The set screw used to lock the length in place is highlighted by the blue arrow.}
\end{figure}

The quartz rod is then fed through the flask connector which has been trimmed from the SNO+ water phase laserball design in order to reduce self-shadowing. The central bore has a diameter of 4.2~mm to limit contact between the quartz rod and stainless steel. This section also provides fine tuning of the light injection point by adjusting the length of the screw fit between the source and flask connectors, as highlighted in figure~\ref{fig:adjustment}. As detailed in sections~\ref{subsec:simisotropy}~and~\ref{subsec:isotropyResults}, the light injection height is critical to the polar light intensity distribution uniformity. With the quartz rod fixed in place by the PTFE clamp, (un)screwing the connection effectively (raises) lowers the light injection point within the flask. The screw length can be adjusted by approximately 9~mm total range, or $\pm~4.5$~mm dynamic range. The quartz rod is therefore measured and cut to a 5~mm accuracy for the optimum light injection point. The connector is then adjusted to provide fine tuning. This mechanism also allows adjustments over time to account for potential drift. 

\begin{figure}[htbp]
\centering 
\includegraphics[width=.4\textwidth,origin=c]{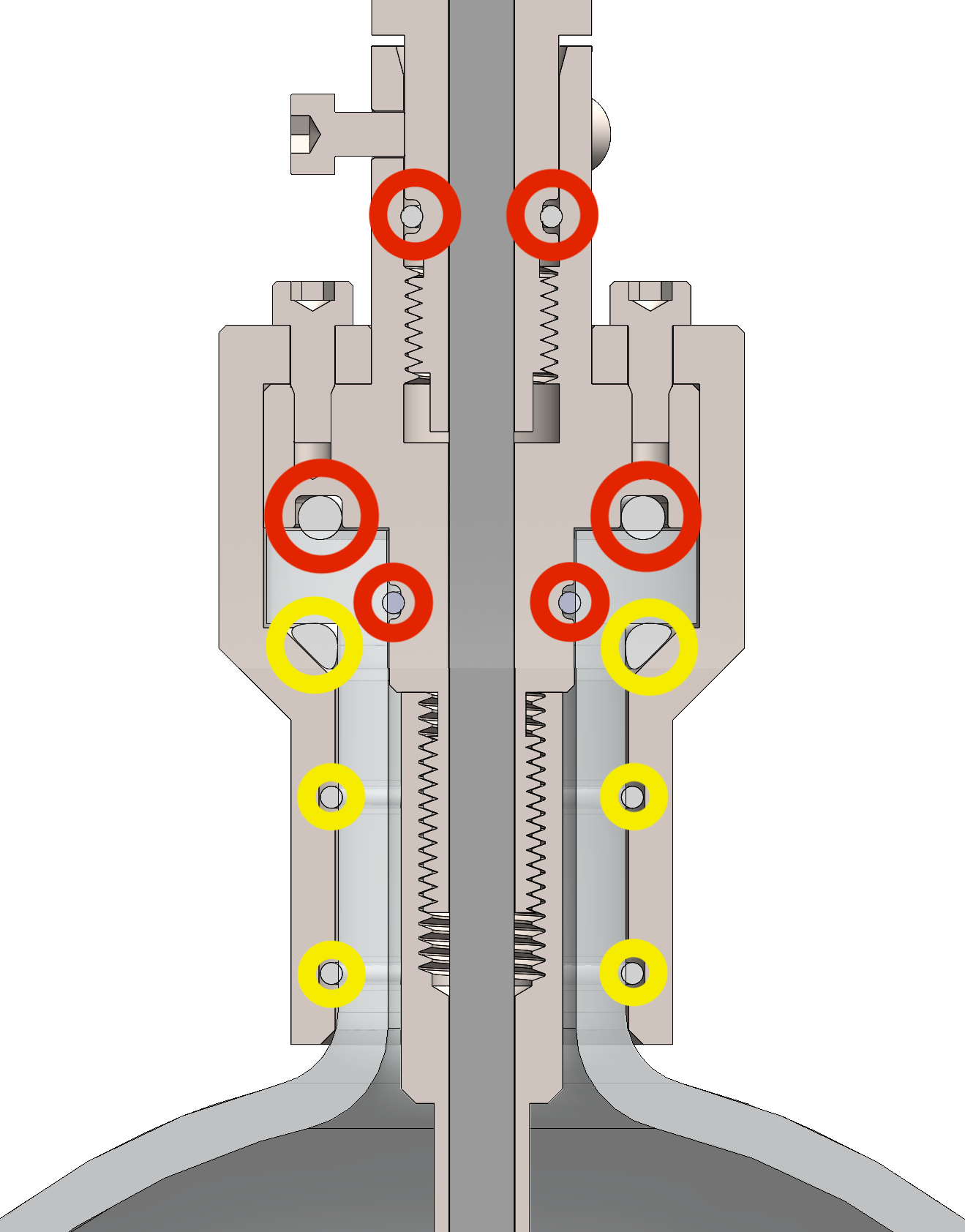}
\caption{\label{fig:orings} The neck connector used to attach the stainless steel assembly to the diffusing flask. The liquid scintillator proofing o-ring gaskets are highlighted by red circles which isolate the internal assembly from the LAB. Yellow circles represent o-ring gaskets used to cushion the quartz flask from the stainless steel.}
\end{figure}

The stainless steel assembly is attached to the diffusing sphere using a neck connector, shown in figure~\ref{fig:orings}. A number of FFKM o-ring gaskets are used for two distinct primary purposes: three gaskets are used to isolate the liquid scintillator from the internal components of the laserball assembly; three additional gaskets are used to cushion the quartz diffusing flask from the neck clamp. Multiple gaskets are used throughout the assembly providing redundancy measures.

\subsection{Diffusing sphere}

Downstream from the assembly the quartz rod is injected into the diffusing sphere, which provides the quasi-isotropic light source needed for detector calibration. The diffusing sphere consists of a ($109\pm1$)~mm diameter spherical quartz flask, with a ($38\pm1$)~mm long neck extension. The diffusing material consists of hollow glass microspheres (20--70~\textmu m diameter, 1~\textmu m thickness) suspended homogeneously in 600~mL of silicone gel. The concentration of the glass microspheres is directly correlated to the scattering properties of the diffuser. The concentration is optimised using the studies presented in this paper to achieve a quasi-isotropic light distribution with low temporal dispersion.

\section{Simulations}
\label{sec:sim}

\begin{figure}[htbp]
\centering 
\includegraphics[width=0.9\textwidth,origin=c]{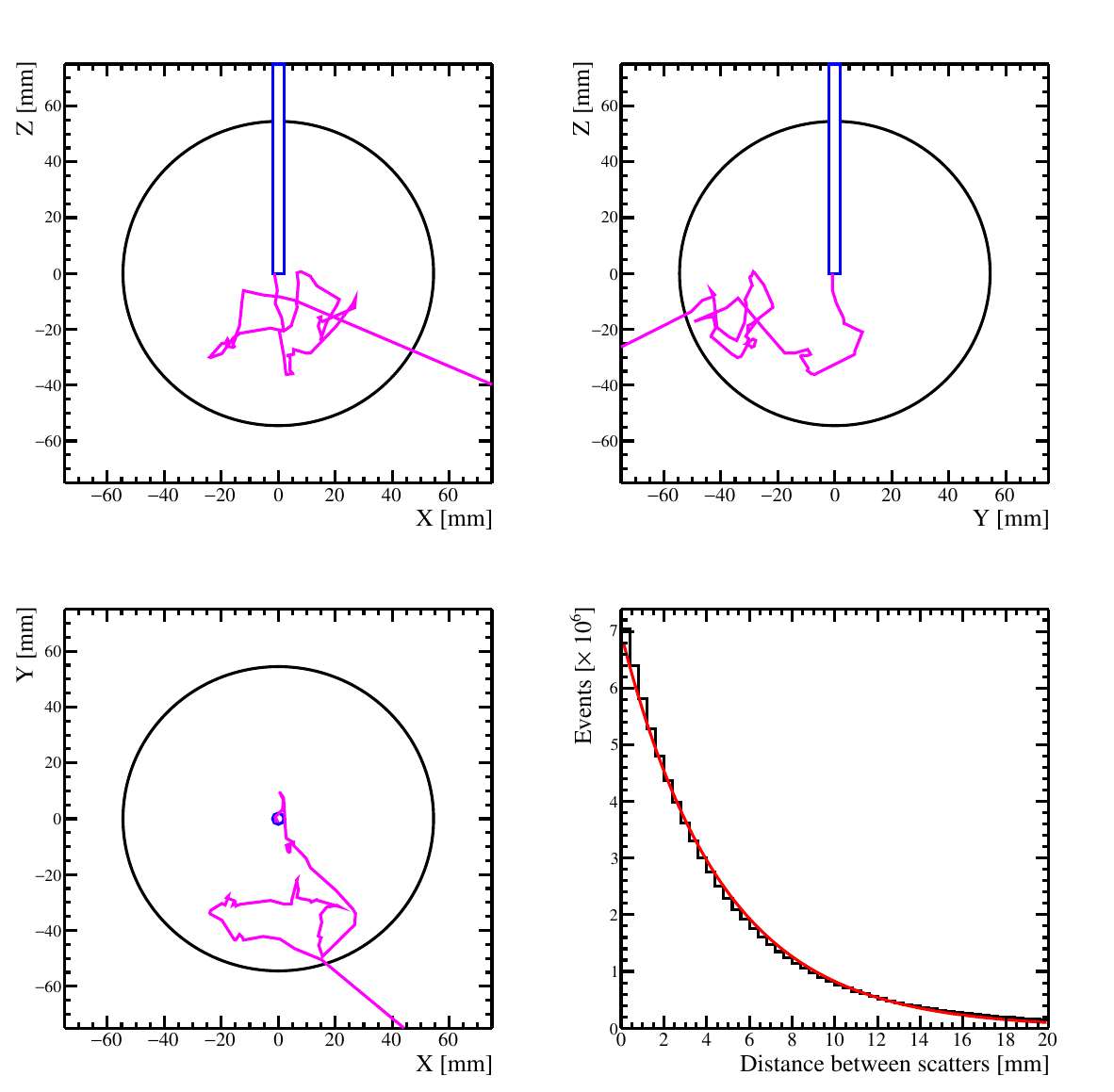}
\caption{\label{fig:photontrack} An example single photon track escaping the diffuser flask for a 2~mg/mL glass microsphere concentration laserball; seen from the front (top left), side (top right), and top (bottom left). The flask wall is shown in black, the quartz rod in blue, and the photon track in pink. The statistical distribution for the distance between scatters over $10^{6}$~photons is also shown (bottom right). An exponential fit (red) is applied to calculate a scattering length of 3.7~mm.}
\end{figure}

A full Monte-Carlo (MC) simulation\footnote{Laserball-MC repository: \url{https://github.com/svalder/laserball-MC/}.} of the diffusing flask is used to model the optical and temporal properties of the laserball. A photon is injected into the centre of a sphere filled with an optically transparent medium with a scattering length, $\lambda$, calculated from 
\begin{equation}
    \lambda = \frac{4\rho{(r^{3} - (r-d)^{3})}}{3Cr^{2}}~,
\end{equation}
where $C$, $r$, $\rho$, and $d$ are all input parameters defining the glass microsphere concentration, radius, density, and wall thickness respectively~\cite{rfordmasters}. The initial position is randomly generated within the exit area of the quartz rod used to inject the light. The initial azimuthal and polar direction is randomly assigned a value between $0 \leftrightarrow 2\pi$, and $0 \leftrightarrow N_A$ respectively, where $N_A$ is the numerical aperture of the optical fibre. The photon then travels a distance $-\lambda\log(x)$ where $x$ is a random variable in the range $0\leftrightarrow1$. If the distance travelled is less than the distance to the flask wall, the photon will propagate this distance and then scatter. Otherwise, the photon will propagate up to the flask wall and then either reflect back into the medium or refract through the glass and into a chosen external medium. Once the photon has exited the diffusing flask, it propagates in a straight line until it is detected. Figure~\ref{fig:photontrack} provides an example simulated single photon track through the diffuser flask. 

\subsection{Scattering} \label{subsec:scaterring}

Whenever a photon scatters off a glass microsphere, the simulation considers three modes based on an approximated impact parameter defined as $b = n_g\sqrt{x}$, where $n_g$ is the refractive index of glass:
\begin{enumerate}[label=(\alph*)]
    \item For forward scattering on the glass microsphere (when $b > 1$), the polar scattering angle is randomly distributed using the following expression
    \begin{equation}
        \cos{(\theta)} = 2x - 1~.
    \end{equation}
    \item If the photon scatters through the hollow glass microsphere (when $b \leq 1$), refraction is taken into account at all interfaces. The scattering angle becomes
    \begin{equation}
        \cos{(\theta)} = 2 \left(n_sx + \sqrt{(1-x)(1-n^2_sx)} \right)^{2} -1~,
    \end{equation}
    where $n_s$ is the refractive index of silicone gel. 
    \item There is a probability of reflecting off the glass microsphere despite $b \leq 1$, in which case the scattering angle is defined as in mode (a). This probability is obtained by averaging over the Fresnel coefficients for the transition between the two materials. \label{item:scatteringC}
\end{enumerate}

\subsection{Reflection} \label{subsec:reflection}

When a photon reaches the interface at the edge of the flask it can either reflect inwards or refract outwards depending on the following parameter
\begin{equation}
    \sin{(\Omega)} = n^2_g (1-|\cos{(\alpha)}|)~,
\end{equation}
where $n_g$ is the refractive index of glass and $\alpha$ is the angle between the photon direction and the normal to the flask surface. From here the photon can undergo three possible modes:
\begin{enumerate}[label=(\alph*)]
    \item If $\sin{(\Omega)} > 1$, the photon will reflect internally with the outgoing angle equal to the incident angle.
    \item If $\sin{(\Omega)} \leq 1$ the photon may still reflect with the same probability outlined in section~\ref{subsec:scaterring} mode \ref{item:scatteringC}, but assuming a transition between the silicone gel and glass.
    \item There is a probability reflections will not occur based on the averaged Fresnel coefficients for the transition between glass and silicone. In this case the photon will refract according to section~\ref{subsec:refraction}.
\end{enumerate}

\subsection{Refraction} \label{subsec:refraction}

When a refraction occurs at an interface between two mediums, the coordinate frame is rotated into a new simplified frame where the $x$-$z$~plane is aligned to the direction of the incident photon normal to the flask surface. Within this 2D frame, $y=0$ and the direction of the refracted photon is calculated using Snell's law. The new direction is then rotated back into the original frame. Refraction is not yet treated between the wall of the flask and the external medium.

\subsection{Isotropy} \label{subsec:simisotropy}

A laserball design specification for calibration is a near-uniform isotropic light output. The simulated light out as a function of azimuthal angle is uniform in all cases; however there is a significant anisotropy in polar angle inheriting from a bias in the direction of light injection. The polar distribution can be modified by tuning the amount of scatters before exiting the flask either by changing the scattering medium, or by displacing the light injection point vertically. 

\subsubsection{Glass microsphere concentration}

\begin{figure}[htbp]
\centering 
\includegraphics[width=.7\textwidth, trim=0 0 0 30,clip]{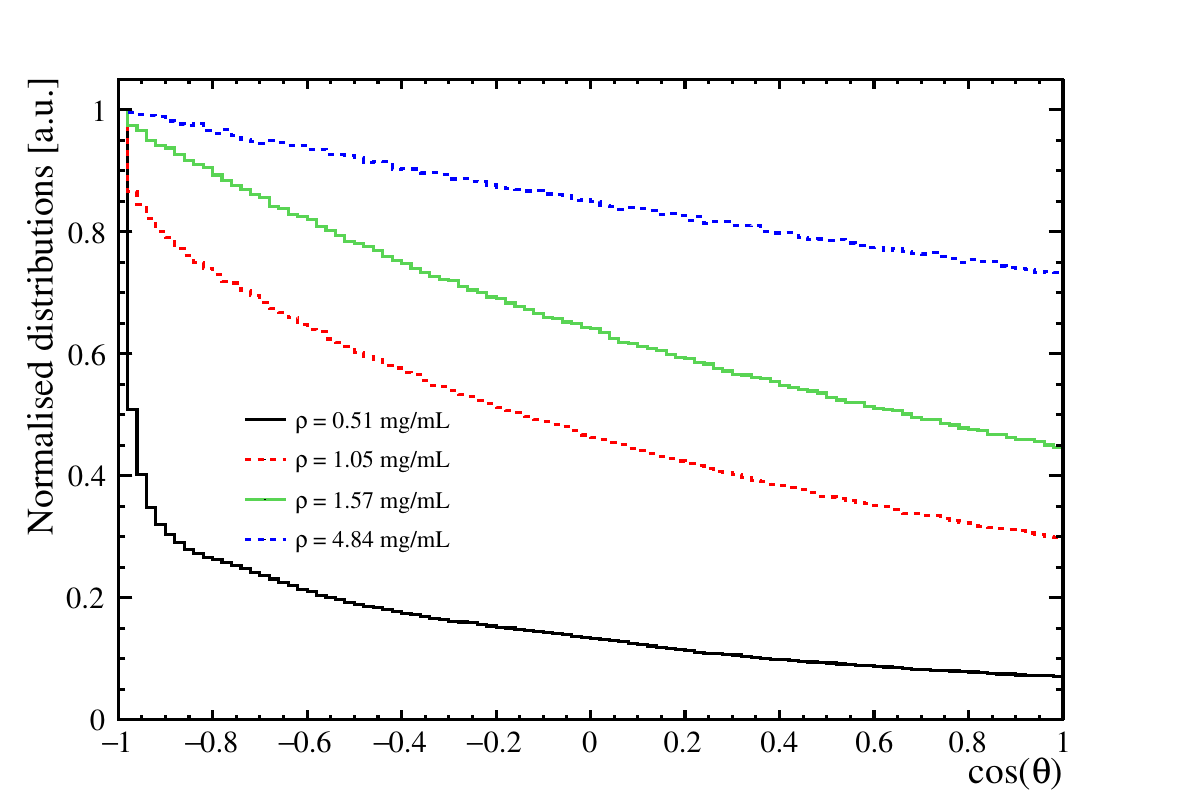}
\caption{\label{fig:simIsotropy} Normalised light intensity output of simulated laserballs sampled with $10^7$ photons as a function of polar angle. Different glass microsphere concentrations have been simulated to best represent tested laserballs under laboratory conditions. The light injection point is at the centre of the laserball.}
\end{figure}

Figure~\ref{fig:simIsotropy} shows the light output uniformity as a function of polar angle for different glass microsphere concentrations, $\rho$. Each plot is normalised so the maximal light intensity is unity. As the number of scatters increases with glass microsphere concentration, the uniformity improves. The profile for a 0.5~mg/mL laserball is heavily biased towards the direction of light injection and appears beam-like. As the glass microsphere concentration increases the polar profile becomes increasingly linear and uniform. Nevertheless, even the largest concentration laserball cannot provide pseudo-isotropy to the level of 10\% desired for SNO+ calibration. Moreover the glass microsphere concentration has a positive correlation with temporal dispersion (section~\ref{subsec:simTiming}), worsening the calibration performance. The solution is to alter the light injection point.

\subsubsection{Light injection point} \label{subsec:simLI}

\begin{figure}[htbp]
\centering 
\includegraphics[width=.47\textwidth, trim=0 0 0 30,clip]{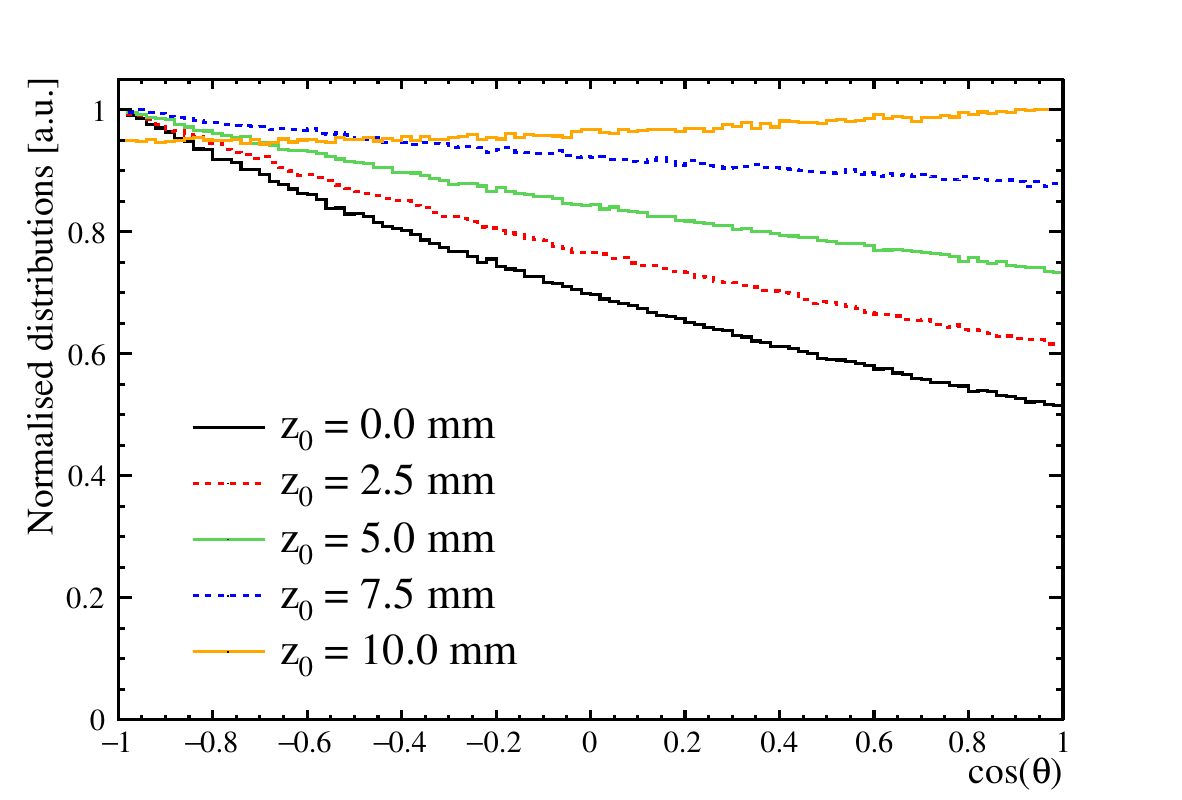}
\qquad
\includegraphics[width=.47\textwidth, trim=0 0 0 30,clip]{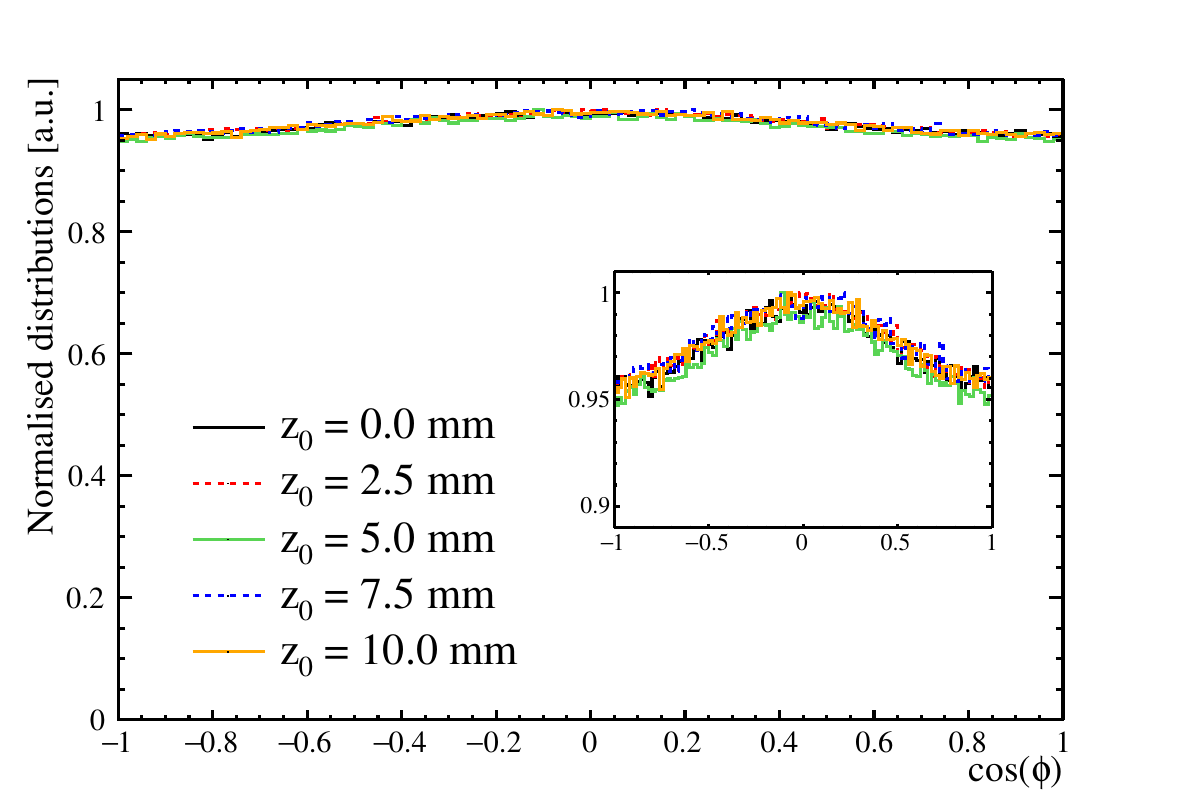}
\caption{\label{fig:simInjectionIsotropy} The normalised light intensity of a 2~mg/mL laserball as a function of (left) polar angle, and (right) azimuthal angle. A number of different light injection displacements, z$_0$, upstream with respect to the center of the laserball have been studied.}
\end{figure}

The inherent forward going bias in polar anisotropy originates from the direction of light injection (defined parallel to the negative z-axis). By displacing the injection point in the axis of injection, it is possible to tune the polar isotropy, without interfering with the azimuthal uniformity. The relationship between z-axis light injection displacement and angular light distribution is shown in figure~\ref{fig:simInjectionIsotropy}. A direct correlation is shown between the shape of the polar light profile and the z-axis light injection point. By displacing the injection point upstream of the laserball center the polar profile shifts favouring photons exiting the flask towards the back of the diffuser. This provides evidence that a combination of the glass microsphere concentration and light injection point can be used to tune a quasi-isotropic light source in both azimuth and polar directions. Nevertheless, the final parameters will also need to be tuned with respect to temporal performance. 

\subsection{Timing} \label{subsec:simTiming}

The temporal performance of the laserball is defined using the escape time from the flask. This is calculated by dividing the total track length inside the flask by the velocity of light within silicone gel. For a timing calibration device, the timing spread needs to be significantly below the temporal resolution of the PMTs. For SNO+, this requirement is safely set to below 1~ns, see section~\ref{sec:calibrequirements}. An investigation into the effect of changing the glass microsphere concentration in the scattering medium on the temporal resolution is outlined below. In addition, a study on the timing distribution as a function of angle is presented, including the dependence on the light injection point. 

\subsubsection{Glass microsphere concentration}

\begin{figure}[htbp]
\centering 
\includegraphics[width=.7\textwidth, trim=0 0 0 30,clip]{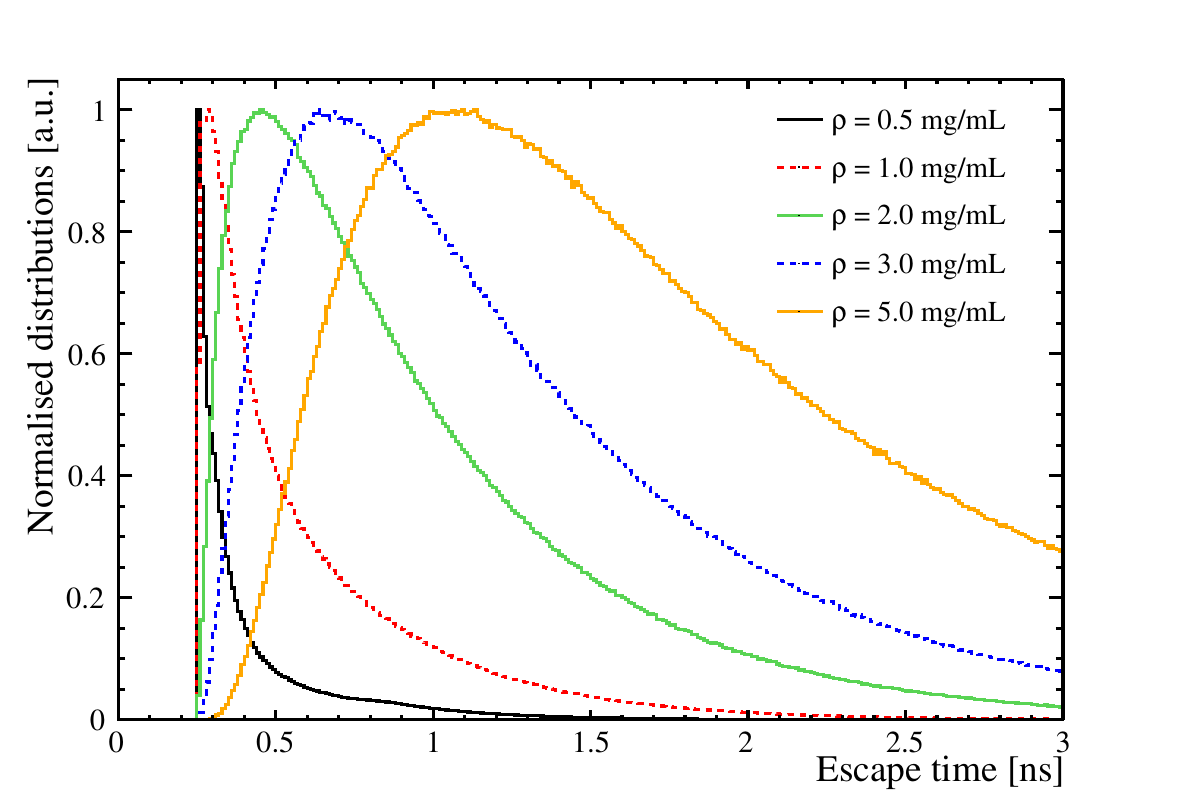}
\caption{\label{fig:simTiming} The normalised average escape times of simulated laserballs sampled with $10^7$ photons. Different glass microsphere concentrations have been simulated to best represent tested laserballs under laboratory conditions.}
\end{figure}

Laserballs with five different glass microsphere concentrations ranging from 0.5~mg/mL to 5.0~mg/mL have been simulated. The timing profiles of simulated laserballs as a function of the glass microsphere concentration, $\rho$, are shown in figure~\ref{fig:simTiming}. The timing spread of each laserball is evaluated using the full-width-half-maximum (FWHM) of the profile. As the glass microsphere concentration increases, the scattering length decreases. As a result, the mean escape time lengthens and timing profile is smeared. A simulated glass microsphere concentration of up to 2~mg/mL seems to best match the SNO+ PMT timing calibration requirements. At 2~mg/mL the simulated pulse width is calculated to be 0.71~ns, providing a comfortable 0.3~ns contingency window. At 5~mg/mL the pulse width increases to 1.64~ns; for 1~mg/mL the pulse width is 0.20~ns, however there is a trade-off in light distribution uniformity as outlined in section~\ref{subsec:simLI}.


\subsubsection{Light injection point}

Simulated laserballs were also produced with different light injection points relative to the centre of the diffusing flask. Five light injection points were simulated, displacing the quartz rod upstream from the centre of the flask between 0~mm and 10~mm. A glass microspheres concentration of 1.57~mg/mL was used enabling direct comparisons with laserball prototypes to be made in section~\ref{subsec:timingResults}. When averaging over the escape times of all photons exiting the flask, no change was seen as a function of light injection point. 

\begin{figure}[htbp]
\centering 
\includegraphics[width=.7\textwidth, trim=0 0 0 30,clip]{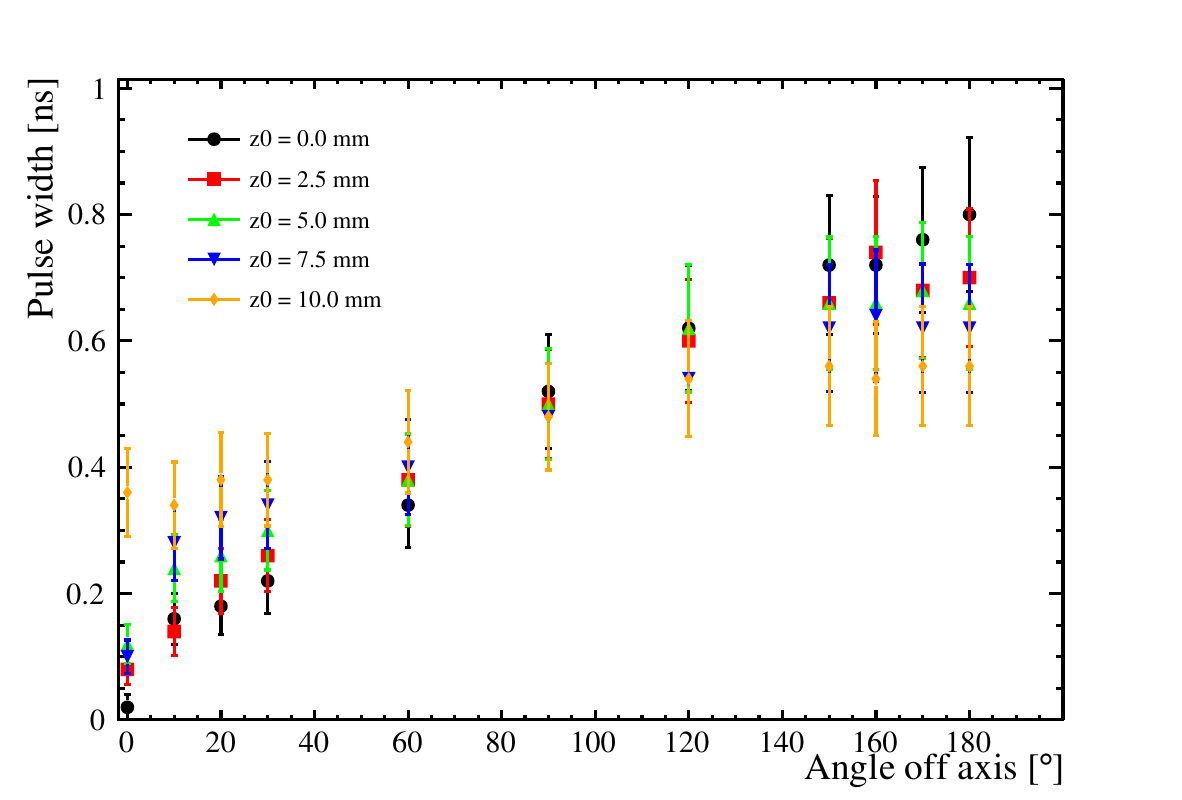}
\caption{\label{fig:simTimingAngle} The timing profiles detected at different polar angles for a 1.57~mg/mL glass microsphere concentration laserball. The study has been performed over multiple light injection points, $z_0$, displacing the quartz rod further upstream. On-axis is defined as the light injection direction where $0^\circ$ ($180^\circ$) is forward (backwards) going. The simulation was produced with $10^8$ photons.}
\end{figure}

Further simulations were also performed in order to investigate the timing resolution as a function of angle. Instead of averaging the escape time over all simulated photons, the photons are captured by virtual detectors placed at regular polar angle intervals from $0^\circ$ (forward-going) to $180^\circ$ (backwards-going). The average escape time is then calculated for captured photos at each virtual detector independently. The virtual detectors are designed to be $8$~mm~$\times~8$~mm~$\times~8$~mm and placed 118~mm from the laserball centre to best represent the laboratory conditions outlined in section~\ref{subsec:timingSetup}. The results are shown in figure~\ref{fig:simTimingAngle}. For large off-axis polar angles the timing profile increases linearly. The gradient of the increase was noted to decrease as a function of light injection point displacement, indicating that the timing profile uniformity can be improved by displacing the quartz rod. Nevertheless, all results for a 1.57~mg/mL glass microsphere concentration are below the 1~ns timing requirements for PMT calibration in SNO+. Uncertainties in calculating the pulse width are derived from the 0.02~ns bin width and the statistical error added in quadrature.

\section{Testing Prototype Performance} \label{sec:tests}

Four prototype laserballs were filled with different glass microsphere concentrations: 0.51~mg/mL, 1.05~mg/mL, 1.57~mg/mL, and 4.84~mg/mL. This section will outline the experimental tests used to evaluate the scattering cross-section of the glass microspheres, the mechanical qualities of the laserball assembly, and the temporal and optical performance of the laserballs as a function of the simulated properties outlined in section~\ref{sec:sim}.

\subsection{Scattering cross-section}

The light transmission through the diffuser mixture must be known for any concentration of the scattering medium. The attenuation factor is expressed as the product of the scattering cross section, $\sigma_S$, and the approximate number density of the glass microspheres, $\rho_N$, from which the mean scattering length, $\hat{x}$, can be derived as
\begin{equation}
    \hat{x} = \frac{1}{\sigma_S\rho_N}~.
\end{equation}
The light transmission was measured by directing light through an optical fibre into a cuvette filled with the scattering medium. A pixel detector measured the transmitted light output from the cuvette. A UV lamp was used as a light source to make measurements from 200~nm to 1100~nm. Quartz cuvettes were filled with silicon gel where the concentration of suspended glass microspheres was varied from 0~mg/mL to 5~mg/mL. A reference spectrum, without glass microspheres, and a dark spectrum, without the lamp, were taken as control measures. From this the optical transmission is calculated using
\begin{equation} \label{eqn:transmission}
    T_{\lambda} = \frac{S_{\lambda} - D_{\lambda}}{R_{\lambda} - D_{\lambda}}~,
\end{equation}
where $S_{\lambda}$, $D_{\lambda}$, and $R_{\lambda}$ are the sampled, dark, and reference intensities at a defined wavelength $\lambda$. The absorbance, $A_{\lambda}$ is also defined as
\begin{equation} \label{eqn:absorbance}
    A_{\lambda} = -\log(T_{\lambda})~.
\end{equation}
Following equations~\eqref{eqn:transmission}~and~\eqref{eqn:absorbance}, the glass microspheres have a working range across the UV-visible spectrum from 300~nm to 800~nm. This spans the entire range of the SNO+ light injection laser. 

\begin{figure}[htbp]
\centering 
\includegraphics[width=.7\textwidth, trim=0 0 0 30,clip]{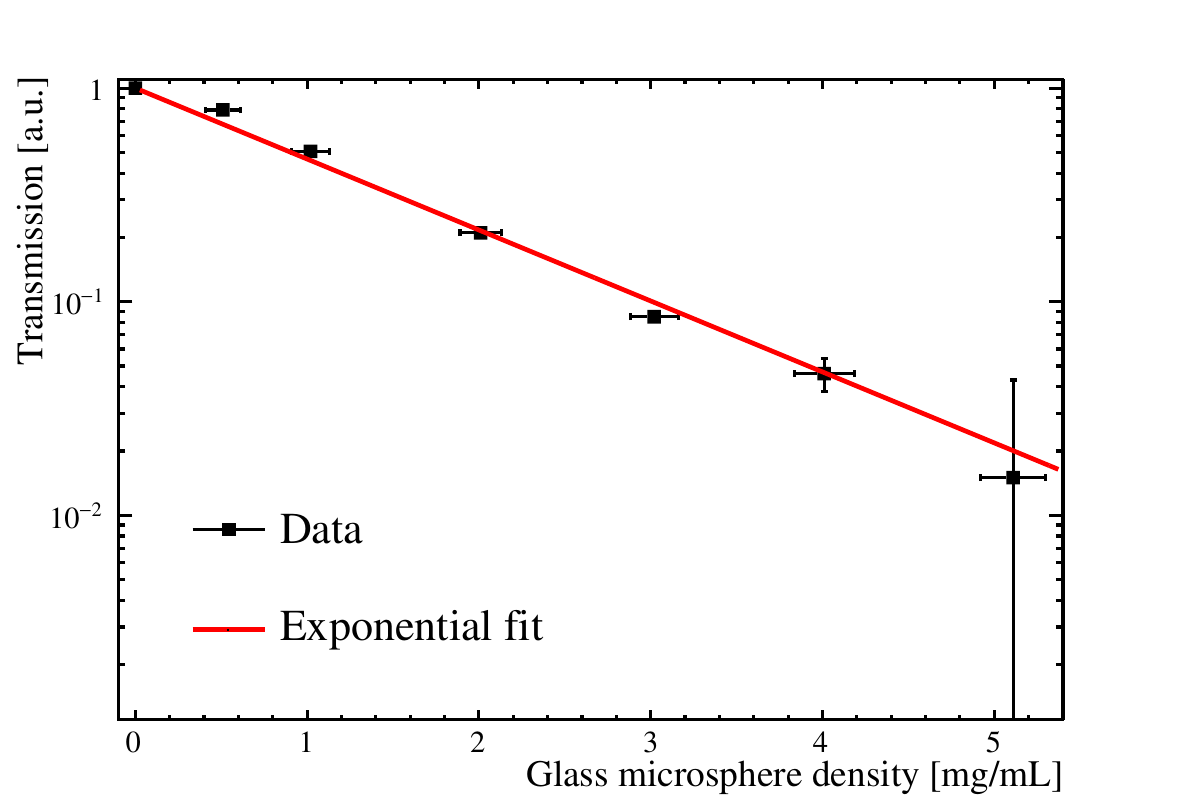}
\caption{\label{fig:cuvette} The transmission in arbitrary units at a wavelength of 400~nm as a function of glass microsphere density. The concentration errors are derived from the precision during mixing, the transmission errors are statistics limited. The cross-section is then evaluated from the slope of a fitted exponential curve (red).}
\end{figure}

The transmission at an example wavelength of 400~nm is shown as a function of glass microsphere density in figure~\ref{fig:cuvette}. The data is fitted with an exponential curve of where the decay constant is extracted to be the scattering cross-section of the glass microspheres. The measured cross-section equates to $(0.765 \pm 0.027)$~cm$^2$/g. This is lower than the cross-section measured for the SNO water phase laserball of $1.35$~cm$^2$/g~\cite{rfordmasters}, and is attributed to the smaller mean particle size of the glass microspheres which is 40~\textmu m in comparison to 50~\textmu m used for SNO. 

\subsubsection{Mechanical tests} \label{subsec:mechanical}

The 4.84~mg/mL concentration laserball flask and assembly were mechanically tested to ensure it functions as designed. At a depth of 12.8~m, equivalent to the bottom of the AV, the maximum hydrostatic pressure the laserball will be under is 1.1~bar. A pressure test was performed up to 4.5~bar over-pressure. The laserball prototype and source connector were submerged in ultra pure water (UPW) and after 20~hours the assembly was opened and found to be dry. The laserball has since passed all pressure tests, with further independent tests in liquid scintillator planned before deployment. 

In addition, the laserball prototype underwent impact tests to test the strength of the flask during the transport and deployment phases. The laserball assembly submerged in UPW was swung from a starting angle as a pendulum against a steel strut. The laserball was found to survive impacts from initial angles up to $30^{\circ}$. At $40^{\circ}$ the impact broke the flask at the top of the neck. The results indicate that the laserball should never hit a solid surface at velocity of more than 1.9~ms$^{-1}$. 

\subsubsection{Radioactivity \& material compatibility tests}

The materials exposed to the external environment consists of 316L stainless steel, quartz, and FFKM o-rings. The SNOLAB low background counting facility~\cite{lawson2011low,lawson2020low} has measured the radioactivity of all materials. The radioactivity of 316L~stainless was evaluated in 2011 via direct $\gamma$-counting using a high-purity germanium detector. It was noticed that the smoothness of the stainless steel surface had a significant effect on the radio-purity, therefore all external (and most internal) pieces have been electropolished. The radio purity of quartz and FFKM o-rings have been measured to ensure they meet the SNO+ radiopurity requirements~\cite{chen2005sno,SNOplus}. Post-manufacturing the quartz flasks were flame treated to smooth the surface and reduce impurities. 

Extensive compatibility and durability studies are conducted over multiple tests to ensure there is negligible chemical effect on either the active medium or the subject material~\cite{anderson2021development, wright2009robust, bartlett2018quality}. Stainless steel 316L and TRP FFKM o-rings, two of the materials in direct contact with the active medium, have been been successfully tested~\cite{SNOplus}. Two TRP FFKM o-rings were soak tested in LAB~+~2~g/l~PPO in 2017. After 1~year of exposure, the material properties remained consistent and no evidence of leaching was found. Further soak tests of all laserball components in liquid scintillator are planned for the near future at Queen's University, Canada.

\subsection{Timing profile} \label{subsec:timingSetup}

\begin{figure}[htbp]
\centering 
\includegraphics[width=\textwidth]{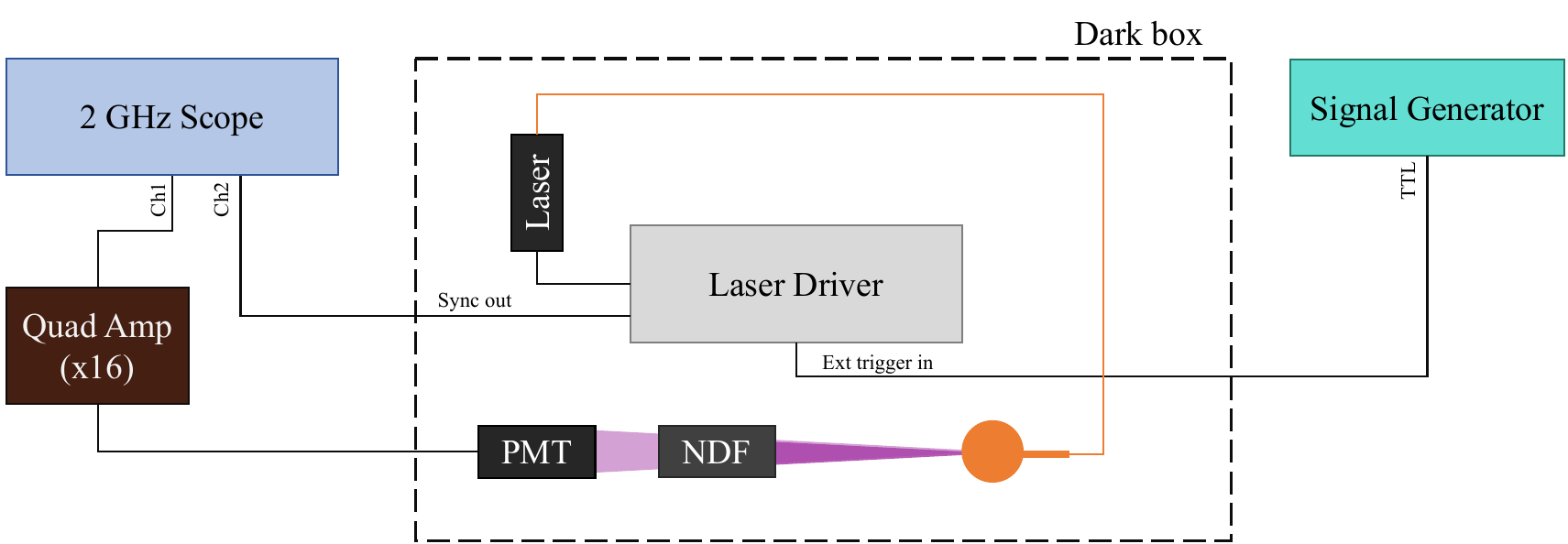}
\caption{\label{fig:timingSetup} A schematic representing the experimental setup used to extract timing measurements. The orange line represents the optical fibre. The orange circle represents the prototype laserball source to be measured.}
\end{figure}

In order to evaluate the temporal performance of the prototype laserballs
an experimental setup was developed at the University of Sussex. The laserball was placed in a dark box and (405~nm) photons were generated by a Picoquant PDL 800-B laser driver. An external signal generator pulsed the laser at 100~kHz with a pulse width of 1~\textmu s. A 2~m long 50~\textmu m core graded index multimode fibre was used to transport the light and was coupled to the quartz rod and injected into the laserball assembly. A PMT was used to detect the light and send a signal downstream via a quad-amplifier ($\times$16) to a fast sampling oscilloscope. In order to achieve a sub-ns resolution, the oscilloscope chosen has a 6.25~GHz sampling rate and 2~GHz bandwidth. A series of neutral density filters (NDF) were used to ensure PMT detects at the single photo-electron (SPE) level. A schematic of the set up is shown in Figure~\ref{fig:timingSetup}. Using the synchronised output from the laser and the output from the PMT, two sequential triggers were set up on the scope. Adjustable delays and trigger timeouts were used to discriminate against potential backgrounds arriving before or after the expected signal. The time separation between the two coincident trigger signals was plotted in a histogram typically over 100,000 acquisitions. Assuming the PMT is in SPE mode, the FWHM of the histogram will define a convolution of the instrument response function (IRF) and the temporal spread caused by the laserball.

\begin{figure}[htbp]
\centering 
\includegraphics[width=.9\textwidth]{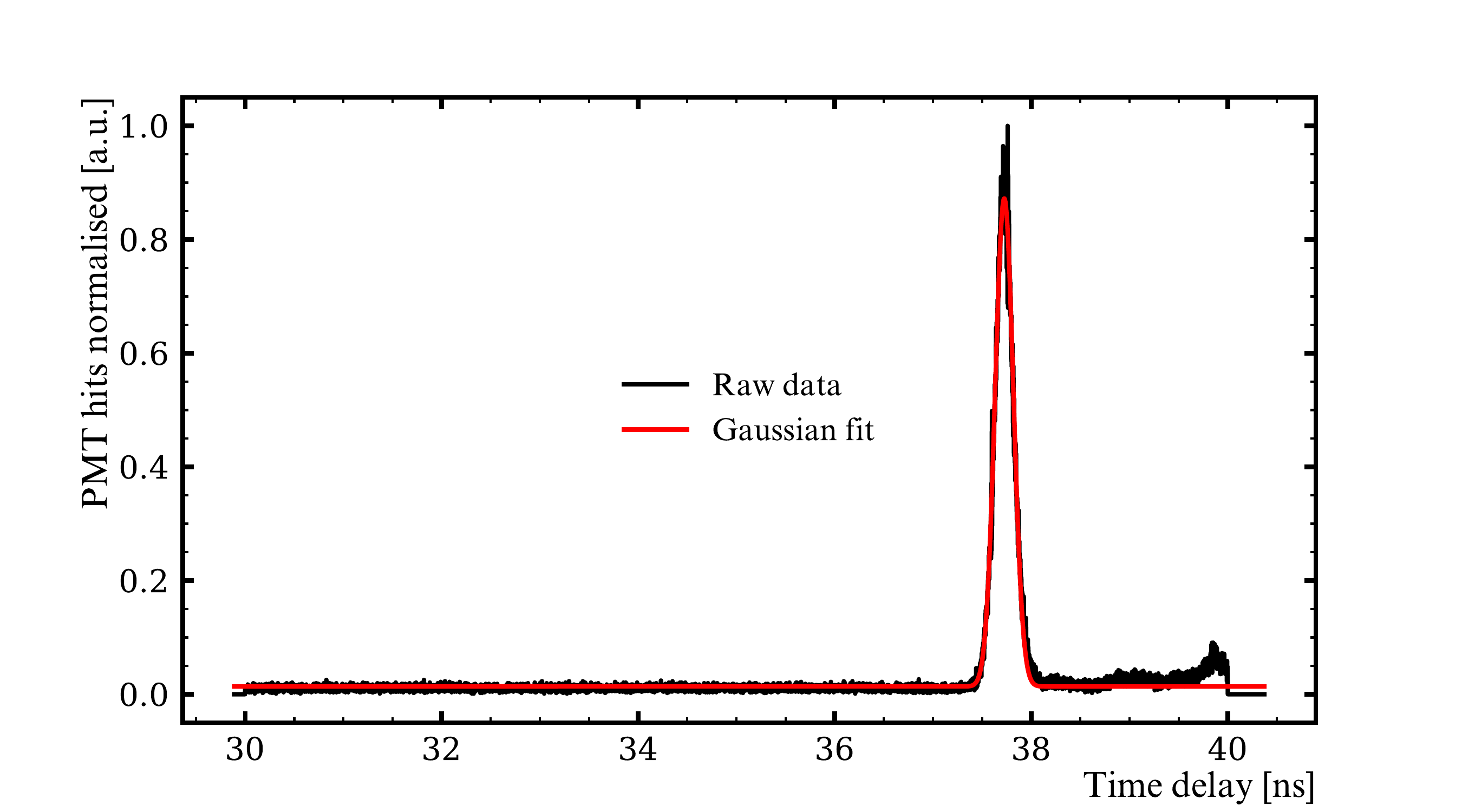}
\caption{\label{fig:scopeTimingMeasurement} The measured time delay between the sequential triggers from the direct light from the laser and optical fibre. The 0.22~ns width of the peak represents the timing profile of the beam and was extracted using a Gaussian fit (red). The secondary peak before 40~ns is attributed to a small optical reflection inside the dark box.}
\end{figure}

To verify sub-ns precision, a validation study measured the timing from a direct laser beam coupled with the optical fibre. Assuming Gaussian distributions, the IRF is a convolution of the various widths of each component added in quadrature:
\begin{equation} \label{eqn:convoluteSetUp}
    \rm{IRF} = \sqrt{(\rm{laser})^2 + (\rm{TTS})^2 + (\rm{fibre})^2 + (\rm{electronics})^2}~.
\end{equation}
The laser has a nominal pulse width of 44~ps which would be seen as a delta function. The dominant systematic contribution comes from the H10721-210 series PMT time transit spread (TTS) and is quoted by the manufacturer as 203~ps. Temporal dispersion from the fibre and electronics are considered to have sub-dominant contributions. Figure~\ref{fig:scopeTimingMeasurement} shows the timing profile of the detected beam over a statistical sample of 100,000 acquisitions. A pulse width of 0.22~ns was extracted from the FWHM of a Gaussian fit. This result is not only in accordance with estimates based on equation~\eqref{eqn:convoluteSetUp}, but also demonstrates sub-ns resolution capability providing a baseline control measurement of the experimental set up.

\subsection{Light intensity isotropy} \label{subsec:isotropySetup}

To evaluate the uniformity of the light emitted from prototype laserballs, raw digital photos were taken and analysed through a custom Python package. The analysis package inputs raw image file and extracts the pixel intensity as a projection across the centre of the laserball in two dimensions. Care is taken to avoid any pre- or post-image processing algorithms. The raw image pixels are converted to a grey scale array, interpreted as a pixel intensity. An algorithm defines the edges of the laserball as the first pixel in a projection above a 10\% threshold of the global maximum pixel intensity. These are used to extract central coordinates of the laserball where horizontal and vertical projections are taken and normalised to the maximum pixel value in the slice. Control photos are taken without the laserball, and with a uniform light source. The control photos are subtracted from the subject image in order to mitigate background noise and independent camera affects respectively.

\begin{figure}[htbp]
\centering 
\includegraphics[width=1.0\textwidth]{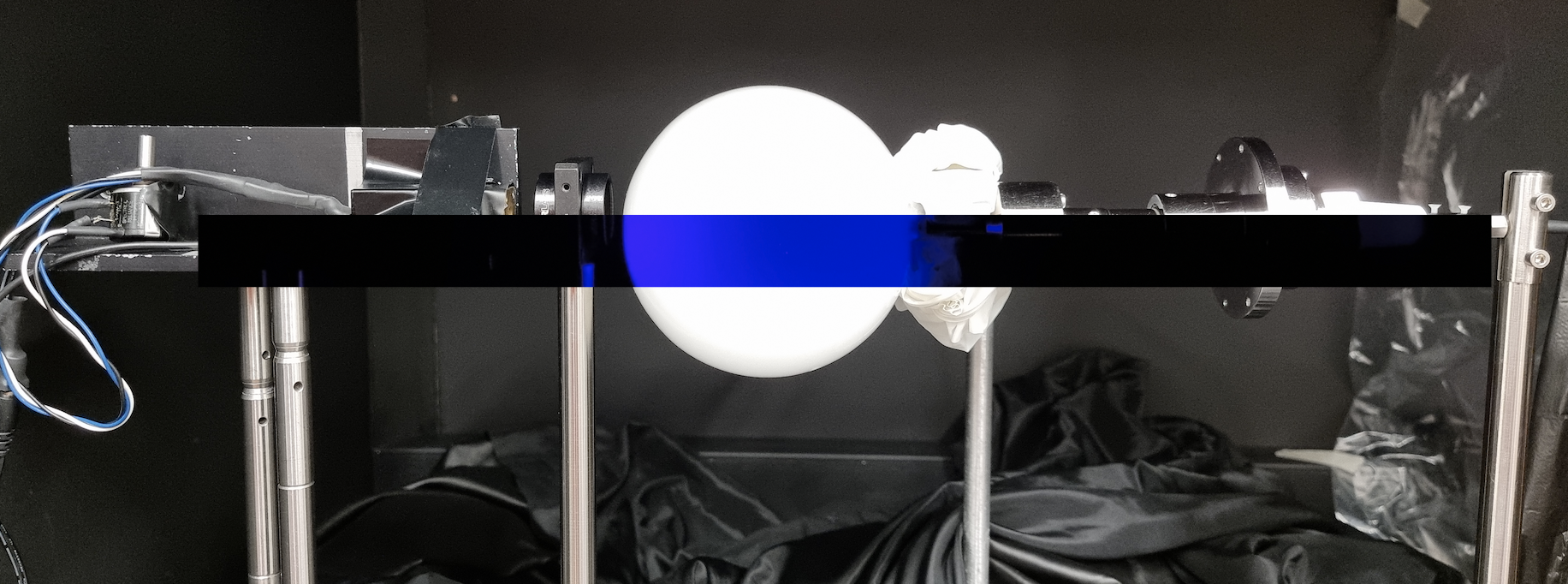}
\caption{\label{fig:overlayedImage} A photograph of the experimental set up used for both timing and light intensity measurements. An overlaid section of the photo can be seen taken in the dark with the laserball on. White gloves have been used to separate the clamp stand and laserball assembly in an attempt to mitigate contamination.}
\end{figure}

The laserball was placed in the dark box and light was injected using the same laser and optical fibre as in section~\ref{subsec:timingSetup}. The pulse signal was generated internally by the laser driver and was pulsed at the maximum frequency (80~MHz) in order to minimise the exposure time needed for each image. The laser driver intensity and camera exposure were tuned through a calibration run to ensure the maximum pixel intensity was not saturated. Photos were taken at a controlled position from the laserball. A photograph of the experimental set up with an overlaid dark image is shown in figure~\ref{fig:overlayedImage}.

\section{Results} \label{sec:results}

This section will summarise the results of the optical and temporal tests outlined in section~\ref{sec:tests}. A number of key independent variables are considered including glass microsphere concentration and light injection displacement which have been compared to simulated results presented in section~\ref{sec:sim}.

\subsection{Timing} \label{subsec:timingResults}

\begin{figure}[htbp]
\centering 
\includegraphics[width=.47\textwidth]{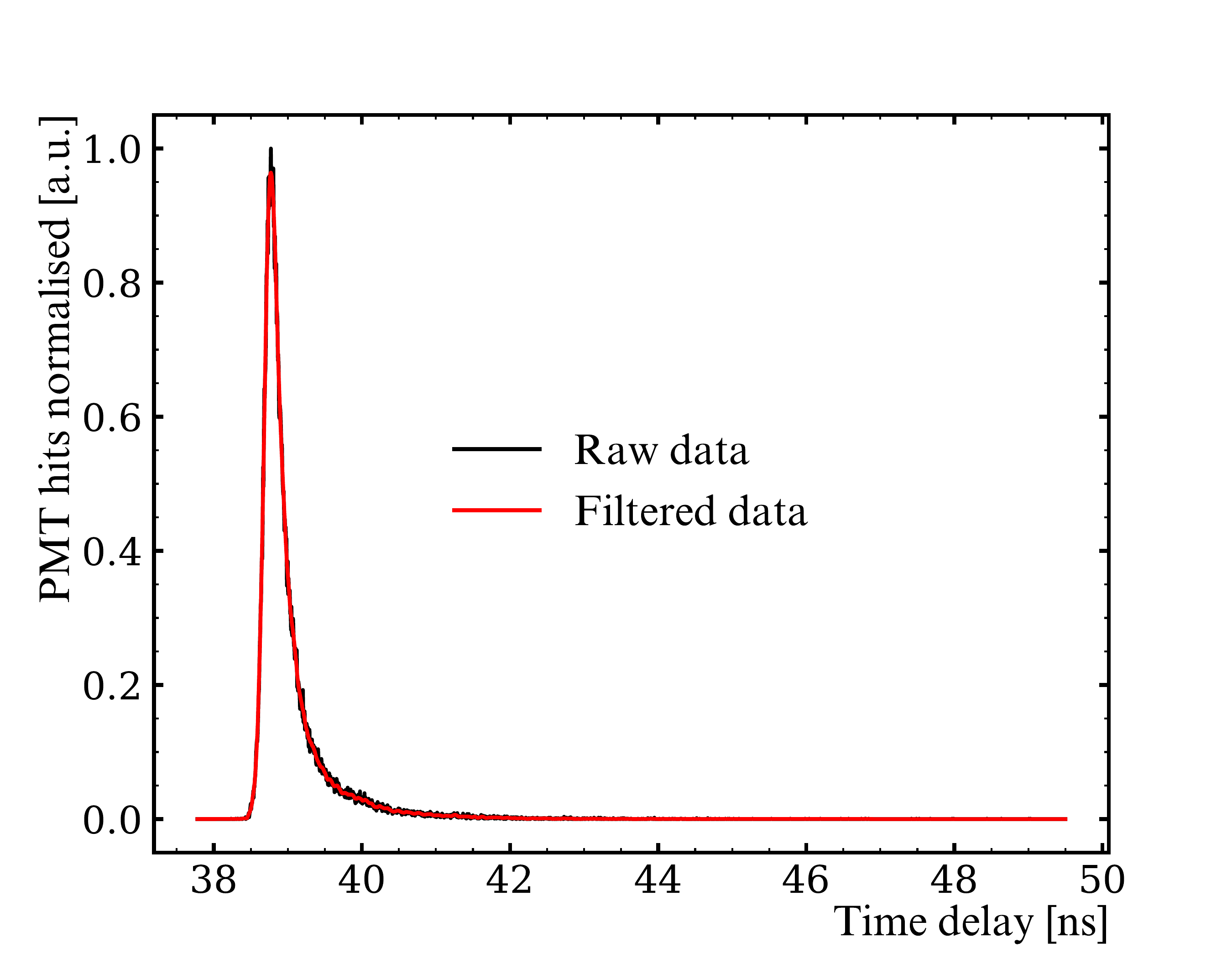}
\qquad
\includegraphics[width=.47\textwidth]{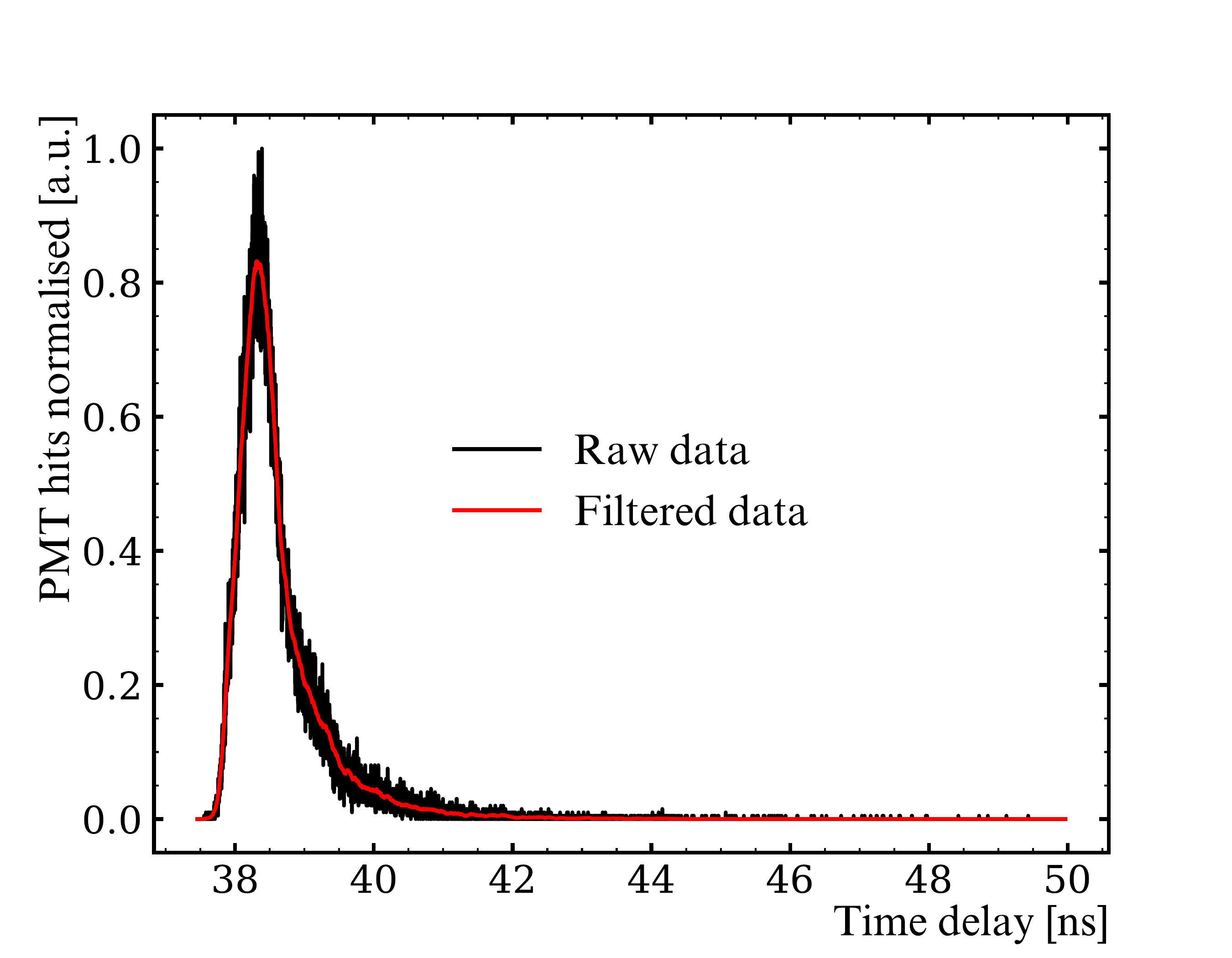}
\caption{\label{fig:timingResults} The on-axis temporal profiles of the 1.05~mg/mL laserball over 100,000 acquisitions (left), and 
the 1.57~mg/mL laserball taken over 20,000 acquisitions (right). The measurement shown is the time delay between the synchronised output trigger from the laser and the PMT trigger. The raw data is shown in black and the filtered data is shown in red.}
\end{figure}

\begin{table}[htbp]
\centering
\caption{\label{tab:timingMeasurements} The temporal resolution for laserballs filled with different glass microsphere concentrations. MC represents the simulated pulse width predicted by section~\ref{subsec:simTiming}; data shows the pulse widths derived from figure~\ref{fig:timingResults}. Data for 0.51~mg/mL and 4.84~mg/mL is not present as they were visibly anisotropic and broken during impact testing respectively, nonetheless they are shown for completeness.}
\smallskip
\begin{tabular}{|c|c|c|}
\hline
Glass microsphere concentration & \multicolumn{2}{|c|}{Timing spread (FWHM) [ns]} \\
\cline{2-3}
[mg/mL] & \hspace*{6.5mm} MC \hspace*{6.5mm} & Data \\
\hline
0.51 & 0.04 & -- \\
1.05 & 0.22 & 0.28 \\
1.57 & 0.51 & 0.64 \\
4.84 & 1.61 & -- \\
\hline
\end{tabular}
\end{table}

A study was performed using the 1.05~mg/mL and 1.57~mg/mL prototype laserballs to evaluate their temporal performance. The 0.51~mg/mL laserball was not selected as it failed the isotropy tests (section~\ref{subsec:isotropyResults}), and the 4.84~mg/mL was not available after the impact tests (section~\ref{subsec:mechanical}). The timing profiles were measured on-axis to the light injection and the resulting profiles can be seen in figure~\ref{fig:timingResults}. The data is normalised to the maximum bin and a Savitzky-Golay filter~\cite{doi:10.1021/ac60214a047} is applied for noise reduction. The pulse width is extracted from the FWHM of the fitted distribution. The results are summarised and compared to simulation in table~\ref{tab:timingMeasurements}. The timing profile increases as a function of glass microsphere concentration and a systematic increase of roughly 20\% is seen between simulation and data. The systematic increase is attributed to the convolution of known experimental effects outlined in equation~\eqref{eqn:convoluteSetUp}, most notably the PMT time transit spread of 203~ps. Importantly, both results are significantly below the 1~ns timing requirements for SNO+ PMT calibration.

\begin{figure}[htbp]
\centering 
\includegraphics[width=.9\textwidth]{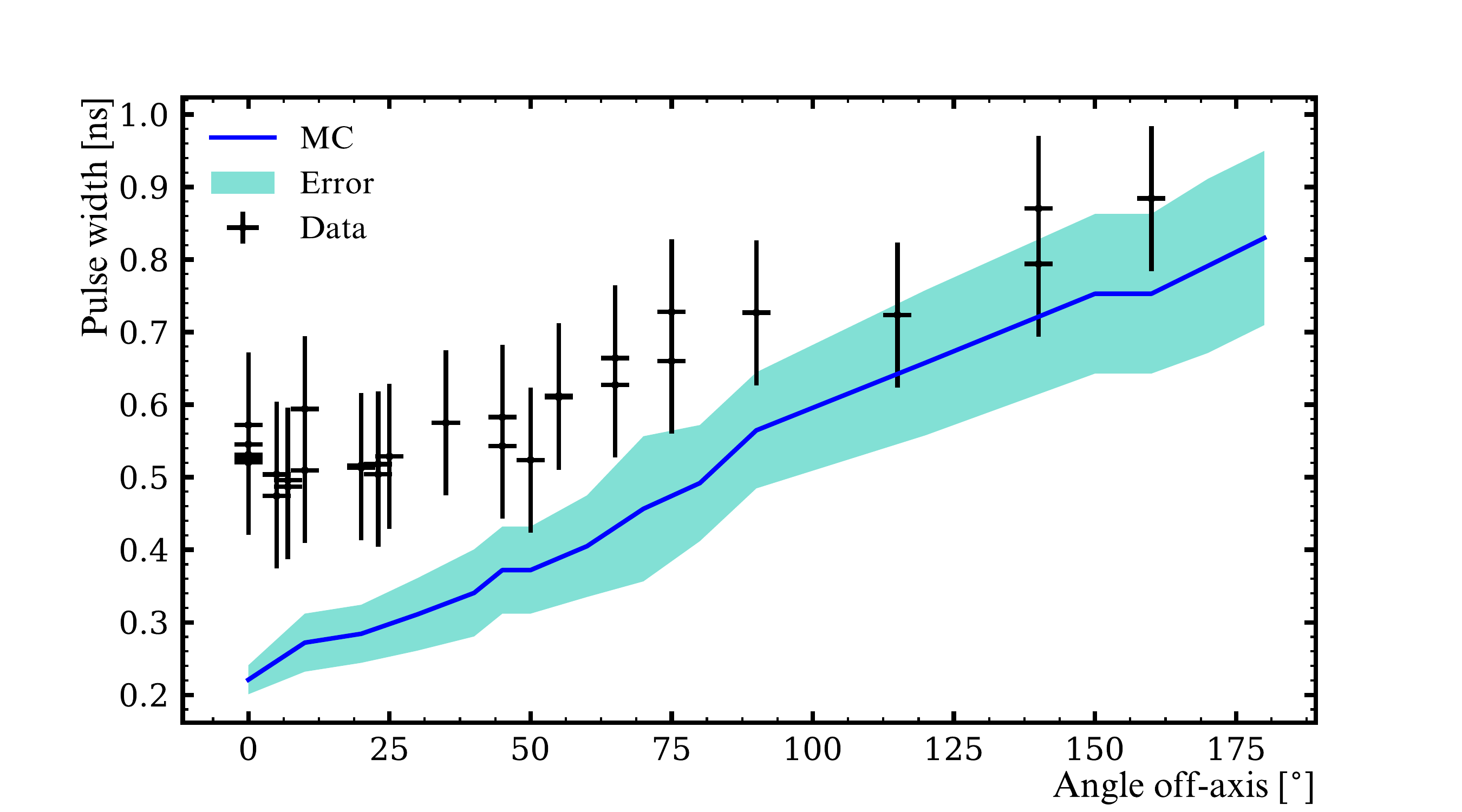}
\caption{\label{fig:timingResultsAngle} The calculated pulse width for a 1.57~mg/mL laserball as a function of polar angle for both data and MC simulations. The simulations are outlined in section~\ref{subsec:simTiming}, and includes an additional 0.22~ns convolution from the pulse width extracted from the control measurement without a laserball, see figure~\ref{fig:scopeTimingMeasurement}. }
\end{figure}

Variations in the temporal resolution were also measured as a function of off-axis angle to the light injection direction. The PMT was moved to off-axis angles ranging from 0$^{\circ}$ to 160$^{\circ}$ where numerous measurements were taken. The results are shown in figure~\ref{fig:timingResultsAngle} where they are also compared to simulations. The dominating uncertainty originates from an experimental timing resolution of 0.2~ns. A positive correlation is seen in the timing as a function of off-axis angle. The simulated measurements include a convolution of the 0.22~ns pulse width extracted from the control measurement in figure~\ref{fig:scopeTimingMeasurement}. There is a discrepancy in temporal uniformity as a function of polar angle between data and simulation. According to figure~\ref{fig:simTimingAngle} temporal uniformity is dependent on the light injection point relative to the centre of the laserball. The discrepancy could therefore be accounted for through a misaligned quartz rod. It should also be noted that the simulations assume a perfect sphere, whereas the laserball flasks are slightly flattened at the bottom which could effect measurements on-axis. Nevertheless, all measured values are below the SNO+ timing calibration requirements of 1~ns. With a resolution above 1~ns, the SNO+ PMTs will be effectively blind to any sub-ns timing variations in angle.  

Independently, the quartz-rod was also displaced by 8~mm, moving the light injection towards the top of the flask. This resulted in an increase in pulse width of 0.08~ns on-axis and 0.10~ns when 30$^{\circ}$ off-axis. This is in line with the MC simulations which predict an increase of 0.08~ns and 0.12~ns respectively as shown in figure~\ref{fig:simTimingAngle}. Any localised spread in timing profile as a function of light injection point is a consequence of the differing timing isotropy and will average out over the full laserball.

\subsection{Isotropy} \label{subsec:isotropyResults}

\begin{figure}[htbp]
\centering 
\includegraphics[width=.471\textwidth, trim=50 40 132 110,clip]{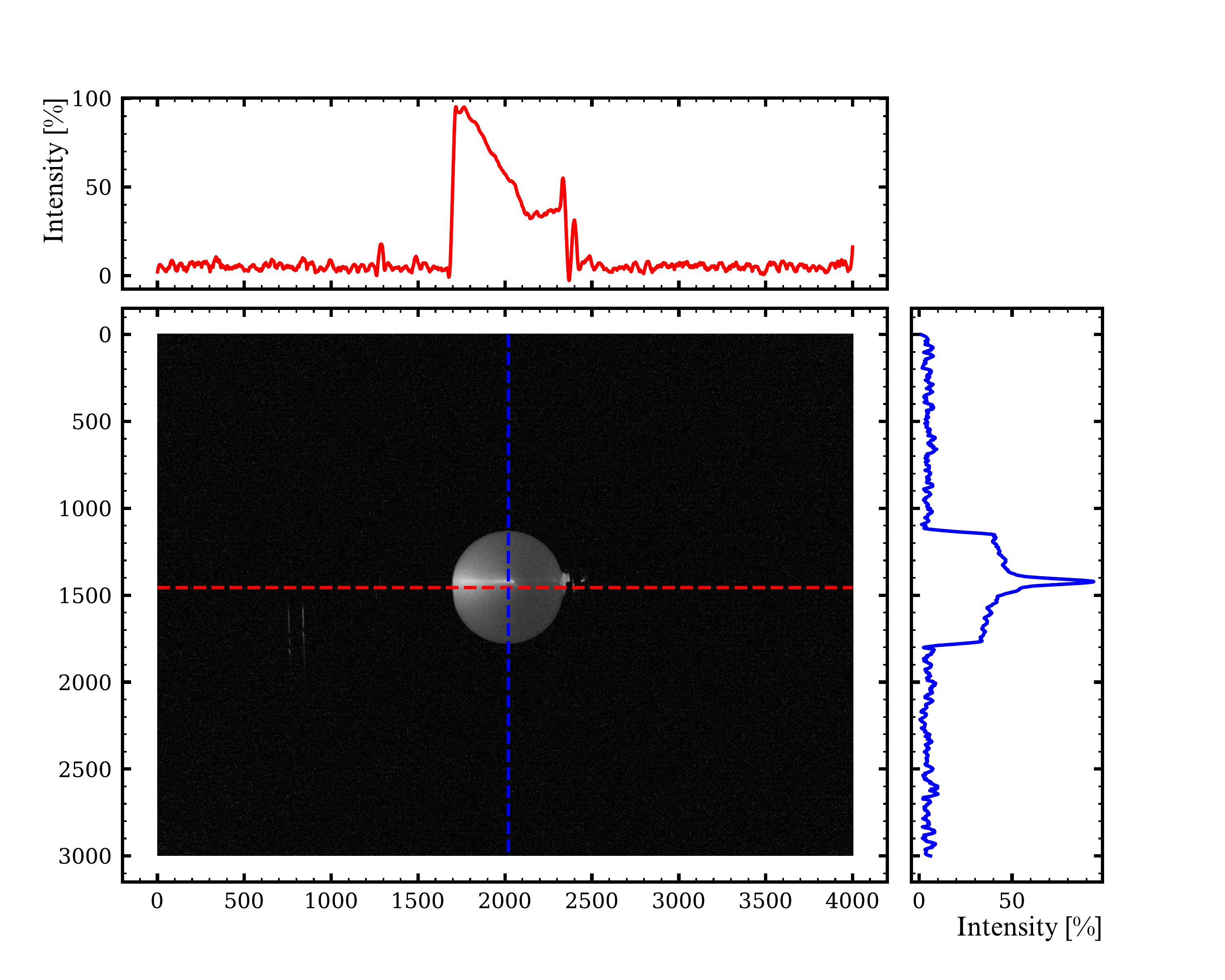}
\qquad
\includegraphics[width=.471\textwidth, trim=50 40 132 80,clip]{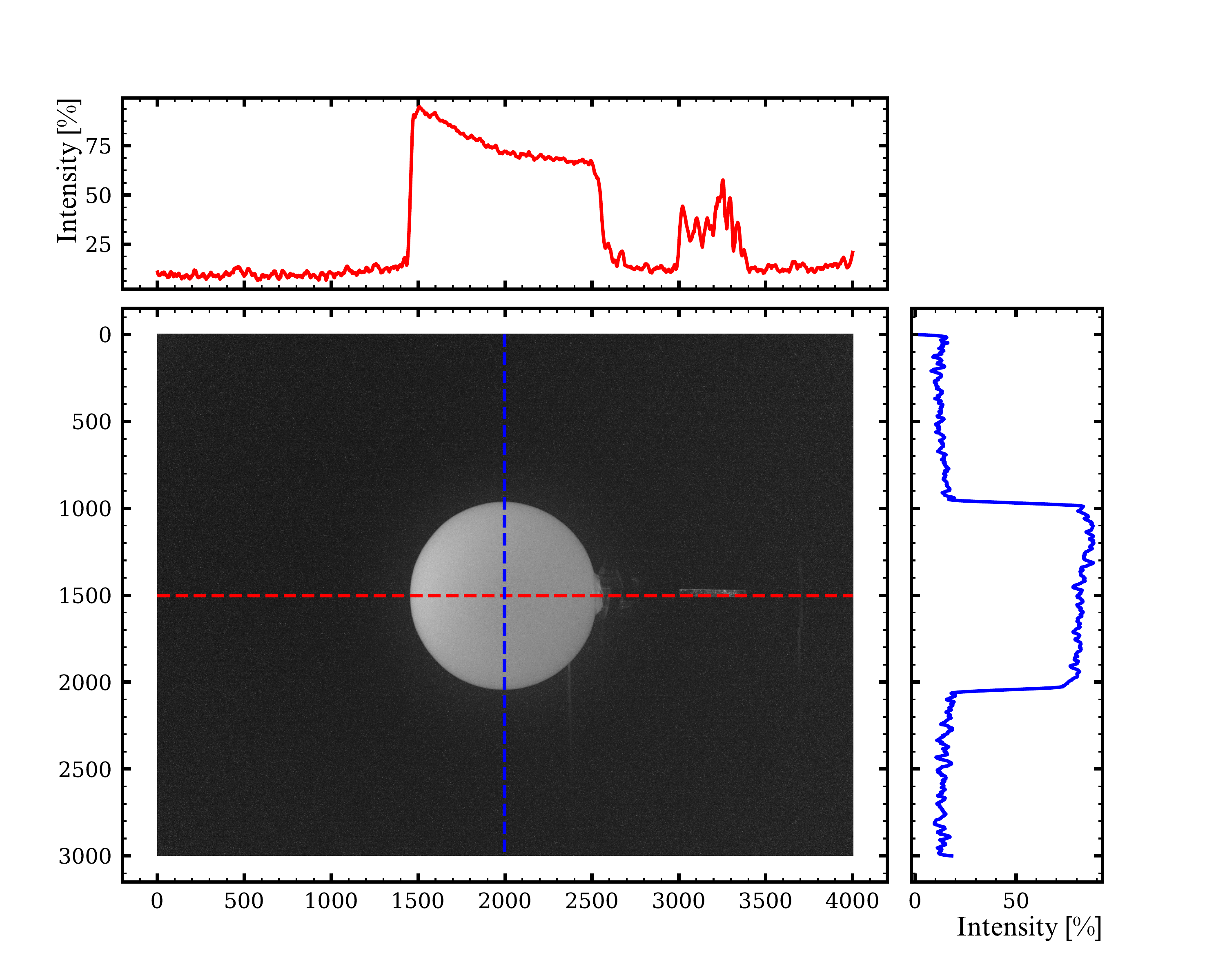}
\caption{\label{fig:isotropyPhotos} Example images of the 0.51~mg/mL (left) and 1.57~mg/mL laserballs (right) respectively. Azimuthal and polar projections are represented by the blue and red lines respectively. Light is being injected from the right hand side of the image. The quartz rod is exposed for the 1.57~mg/mL. }
\end{figure}


Three laserball prototypes with glass microsphere concentrations of 0.51~mg/mL, 1.05~mg/mL, and 1.57~mg/mL were tested for their respective light intensity isotropy performance. Example analysis images of the 0.51~mg/mL and 1.57~mg/mL laserballs are shown in figure~\ref{fig:isotropyPhotos}. It was evident by eye that with a concentration of 0.51~mg/mL, not enough scatters are present to form a near-isotropic distribution, appearing with a more `beam-like' profile. At concentrations of 1.05~mg/mL and 1.57~mg/mL the laserballs act as true diffusers where the light output is nearing isotropic by nature. Nevertheless, a systematic linear polar anisotropy of up to 40\% is seen with the quartz rod terminating at the center of the flask. Given simulations suggest increasing the concentration of diffusing material benefits uniformity and the timing requirements were comfortably met by both prototype laserballs, only the 1.57~mg/mL prototype is considered moving forward.


\begin{figure}[htbp]
\centering 
\includegraphics[width=\textwidth, trim=150 700 200 100,clip]{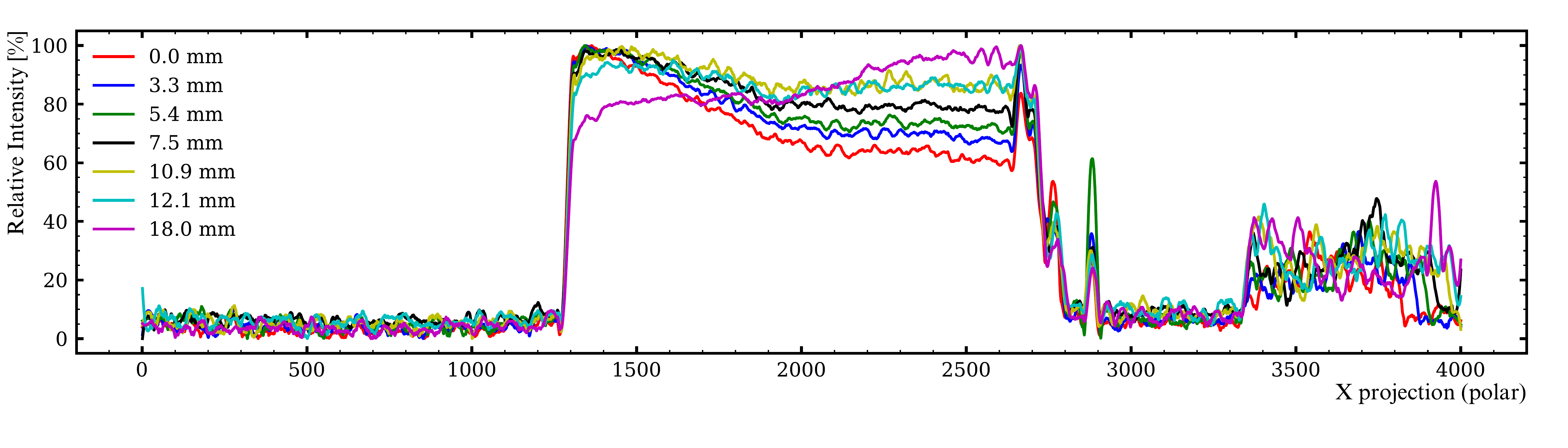}
\qquad
\includegraphics[width=\textwidth, trim=150 50 200 820,clip]{images/anisotropy_data_157_new.pdf}
\caption{\label{fig:isotropyRodDisplacement} Relative isotropy of the 1.57~mg/mL laserball as a function of the light injection point displacement from an initial reference point of 0~mm at the center of the laserball flask. A background subtraction has been applied from a control image. The polar (top) and azimuthal (bottom) distributions are both shown. The peaks at the end of the spectra is from exposed quartz rod where the full assembly has not been used.}
\end{figure}

Simulations suggest that by displacing the light injection point any inherent linear polar anisotropy can be corrected for, see section~\ref{subsec:simLI}. The quartz rod was systematically displaced from a reference point located at the centre of the flask towards the top of the laserball (z-axis). Figure~\ref{fig:isotropyRodDisplacement} summarises the polar isotropy over a number of displacements from 0~mm to 18~mm for a glass microsphere concentration of 1.57~mg/mL. The correlation between light injection point and polar anisoptropy is evident; the further upstream the quartz rod is displaced, the more favoured back-scattered light becomes. This is further backed by the result at 18~mm where the bias has flipped to favour back-scattered light. This result demonstrates the possibility of obtaining a quasi-isotropic polar distribution by tuning the light injection point inside the final laserball. The azimuthal distribution was unaffected by the quartz rod displacement. Both of these results agree with simulation which predicts an isotropic polar distribution at a glass microsphere concentration of 1.57~mg/mL and a light injection displacement of 12~mm.

%


\section{Conclusions}

A new laserball calibration device has been developed for the primary purpose of PMT calibration during the SNO+ scintillator phase of the experiment. Three main design improvements have been implemented to improve calibration performance relative to the water phase laserball. Novel simulations including all key optical properties have been developed to model the propagation of photons within the diffuser flask. Glass microsphere concentration and the light injection point have been identified as the two dominant design characteristics for maximising optical and temporal performance. Experimental and mechanical testing at the University of Sussex has been used to validate the design. Tests on multiple laserballs with different glass microsphere concentrations, over a variety of light injection points, has demonstrated the capability of achieving a quasi-isotropic light distribution with a sub-ns timing resolution matching the calibration requirements. This was achieved using a concentration of 1.57~mg/mL with a vertical light injection displacement of approximately 12~mm upstream. The results presented provide a proof-of-principle for which a final laserball can be developed for optical calibration of the SNO+ detector during scintillator phase. The next steps will be to complete the material compatibility testing with liquid scintillator, construct the final laserball with the desired glass microsphere concentration and light injection point, and fully characterise its light and temporal properties before deployment.


\acknowledgments

We would like to thank the SNO+ collaboration for their fundamental and continued support of this work, providing expertise and feedback at all stages. In particular, we'd like to thank J. Maneira for their advice and guidance throughout the course of this project; as well as C. Kraus, V. Lozza, A. Wright, P. Skensved, and A.S. Inacio for their extensive review of the work. In addition, we'd also like to thank the technical and mechanical staff at the University of Sussex physics department.

This research was supported by the Science and Technology Facilities Council (STFC), United Kingdom, through grants ST/W000512/1 and ST/S003568/1. 


\bibliographystyle{JHEP.bst}

\bibliography{jinst-latex-laserball}













\end{document}